\documentclass[12pt]{article}
\pdfoutput=1
\usepackage{putex}
\usepackage{graphicx}
\usepackage{epstopdf}
\usepackage{enumerate}
\usepackage{cite}
\usepackage{tensor}
\usepackage{slashed}
\usepackage{caption}
\usepackage{subcaption}
\usepackage{bbm}
\usepackage[multiple]{footmisc}
\usepackage[normalem]{ulem}

\numberwithin{equation}{section}

\newcommand{\abs}[1]{\left\lvert #1 \right\rvert}

\newcommand {\be} {\begin {equation}}
\newcommand {\ee} {\end {equation}}

\newcommand {\bes} {\begin {equation*}}
\newcommand {\ees} {\end {equation*}}

\newcommand{\es}[2] {\begin{equation} \label{#1} \begin{split} #2 \end{split} \end{equation}}

\newcommand{\Z}{\mathbb{Z}}

\newcommand{\C}{\mathbb{C}}

\def\Tr{\mop{Tr}}

\newcommand{\beq}{\begin{equation}}
\newcommand{\eeq}{\end{equation}}

\newcommand\ov{\over}
\def\le{\left}
\def\ri{\right}

\def\<{\langle}
\def\>{\rangle}
\newcommand\ket[1]{\ensuremath{\left\lvert{#1}\right\rangle}}
\newcommand\bra[1]{\ensuremath{\left\langle{#1}\right\rvert}}
\newcommand\p{\ensuremath{\partial}}

\newcommand\lam{\ensuremath{\lambda}}

 \newcommand{\K}{ {\widetilde{N_f}}}

 \definecolor{orange}{rgb}{1,0.5,0}

\begin{document}

\preprint{MIT-CTP-4520}

\institution{MIT}{Center for Theoretical Physics, Massachusetts Institute of Technology, Cambridge, MA 02139}

\title{Three-sphere free energy for classical gauge groups}

\authors{M\'ark Mezei and Silviu S.~Pufu}

\abstract{
In this note, we calculate the $S^3$ free energy $F$ of 3-d ${\cal N}\geq 4$ supersymmetric gauge theories with $U(N)$,  $O(N)$, and $USp(2N)$ gauge groups and matter hypermultiplets in the fundamental and two-index tensor representations.  Supersymmetric localization reduces the computation of $F$ to a matrix model that we solve in the large $N$ limit using two different methods.  The first method is a saddle point approximation first introduced in \cite{Herzog:2010hf}, which we extend to next-to-leading order in $1/N$.  The second method generalizes the Fermi gas approach of \cite{Marino:2011eh} to theories with symplectic and orthogonal gauge groups, and yields an expression for $F$ valid to all orders in $1/N$.  In developing the second method, we use a non-trivial generalization of the Cauchy determinant formula.}

\maketitle

\tableofcontents

\section{Introduction}\label{sec:Intro}

In the absence of a perturbative understanding of the fundamental degrees of freedom, one can learn about M-theory only through various dualities.  A promising avenue is to use the AdS/CFT correspondence \cite{Maldacena:1997re,Gubser:1998bc,Witten:1998qj} to extract information about M-theory that takes us beyond its leading (two-derivative) eleven-dimensional supergravity limit. Such progress is enabled by the discovery of 3-d superconformal field theories (SCFTs) dual to backgrounds of M-theory of the form AdS$_4 \times X$ \cite{Aharony:2008ug,Jafferis:2009th,Gaiotto:2009tk,Jafferis:2008qz,Franco:2009sp,Benini:2011cma,Benini:2009qs,Martelli:2009ga}, as well as the development of the technique of supersymmetric localization in these SCFTs \cite{Kapustin:2009kz,Jafferis:2010un,Hama:2010av} (see also \cite{Pestun:2007rz}).  For instance, computations in these SCFTs may impose constraints on the otherwise unknown higher-derivative corrections to the leading supergravity action.

In this paper we study several 3-d SCFTs, with the goal of extracting some information about M-theory on AdS$_4 \times X$ that is not accessible from the two-derivative eleven-dimensional supergravity approximation.  These theories can be engineered by placing a stack of $N$ M2-branes at the tip of a cone over the space $X$.  A good measure of the number of degrees of freedom in these theories, and the quantity we will focus on, is the $S^3$ free energy $F$ defined as minus the logarithm of the $S^3$ partition function, $F = -\log \abs{Z_{S^3}}$ \cite{Jafferis:2011zi,Klebanov:2011gs,Myers:2010tj,Casini:2012ei}.  At large $N$, the $F$-coefficient of an SCFT dual to AdS$_4\times X$ admits an expansion of the form \cite{Drukker:2010nc,Herzog:2010hf} 
 \es{FExpansion}{
  F = f_{3/2} N^{3/2} + f_{1/2} N^{1/2} + \ldots \,.
 }
The coefficient $f_{3/2}$ can be easily computed from  two-derivative 11-d supergravity \cite{Drukker:2010nc,Herzog:2010hf} 
 \es{Gota}{
  f_{3/2} = \sqrt{\frac{2 \pi^6}{27 \Vol(X)}} \,,
 }
whereas the coefficient $f_{1/2}$ together with the higher-order corrections in \eqref{FExpansion} cannot \cite{Bergman:2009zh, Bhattacharyya:2012ye}.  In this paper we will calculate $f_{1/2}$ for various SCFTs with M-theory duals.

We focus on SCFTs with ${\cal N} \geq 4$ supersymmetry.   In such theories, supersymmetric localization reduces the computation of $Z_{S^3}$ to certain matrix models \cite{Kapustin:2010xq}.   For instance, for the ${\cal N} =6$ ABJM theory \cite{Aharony:2008ug}, which is a $U(N)_k \times U(N)_{-k}$ Chern-Simons matter gauge theory, one has \cite{Kapustin:2009kz,Drukker:2010nc}
 \es{ZABJM}{
  Z_{S^3} = \frac{1}{(N!)^2}\int \prod_{i = 1}^N d \lambda_i d\tilde \lambda_i \frac{\prod_{i<j} \sinh^2 (\pi (\lambda_i - \lambda_j)) 
    \sinh^2 (\pi (\tilde \lambda_i - \tilde \lambda_j)) }
    {\prod_{i, j} \cosh^2 (\pi (\lambda_i - \tilde \lambda_j))} \exp \left[i \pi k \sum_i \left(\lambda_i^2 - \tilde \lambda_i^2 \right) \right] \,,
 }
where the integration variables are the eigenvalues of the auxiliary scalar fields in the two ${\cal N} = 2$ vectormultiplets.  This theory corresponds to the case where the internal space $X$ is a freely-acting orbifold of $S^7$, $X = S^7 / \Z_k$. The integral \eqref{ZABJM} can be computed approximately at large $N$ by three methods:
 \begin{enumerate}[I.]
  \item \label{first} By mapping it to the matrix model describing Chern-Simons theory on the Lens space $S^3 / \Z_2$, and using standard matrix model techniques to find the eigenvalue distribution \cite{Drukker:2010nc}.  This method applies at large $N$ and fixed $N/k$.  To extract $f_{3/2}$ and $f_{1/2}$ in \eqref{FExpansion} one needs to expand the result at large 't Hooft coupling $N/k$.
  \item \label{second} By expanding $Z_{S^3}$ directly at large $N$ and fixed $k$ \cite{Herzog:2010hf}.  In this limit, the eigenvalues $\lambda_i$ and $\tilde \lambda_i$ are uniformly distributed along straight lines in the complex plane.
  \item \label{third} By rewriting \eqref{ZABJM} as the partition function of $N$ non-interacting fermions on the real line with a non-standard kinetic term \cite{Marino:2011eh}.  The partition function can then be evaluated at large $N$ and small $k$ using statistical mechanics techniques.
 \end{enumerate}
Using the Fermi gas approach \eqref{third}, for instance, one obtains \cite{Marino:2011eh}
\es{AiryFunc}{
Z={\cal A}(k) \, {\rm Ai}\le[\le({\pi^2 k\ov 2}\ri)^{1/3}\le(N-{k\ov 24}-{1\ov 3k}\ri)\ri]+ {\cal O}\le(e^{-\sqrt{N}}\ri) \ , 
}
where ${\cal A}(k)$ is an $N$-independent constant.  From this expression one can extract
 \es{fABJM}{
  f_{3/2} = k^{1/2} {\sqrt2  \, \pi \ov 3} \,, \qquad
   f_{1/2} = - {\pi \ov \sqrt{2}} \le( {k^{3/2} \ov 24} + {1 \ov 3k^{1/2}} \ri) \,.
 }
These expressions can be reproduced from the first method mentioned above \cite{Drukker:2010nc}, and $f_{3/2}$ can also be computed using the second method\cite{Herzog:2010hf}.

While ABJM theory teaches us about M-theory on AdS$_4 \times (S^7 / \Z_k)$, it would be desirable to calculate $F$ for other SCFTs with M-theory duals, so one may wonder how general the above methods are and/or whether they can be generalized further.  So far, the first method has been generalized to a class of ${\cal N} = 3$ theories obtained by adding fundamental matter to ABJM theory \cite{Santamaria:2010dm}.\footnote{Grassi and Mari\~no informed us that they have also applied the first method to the ${\cal N}=4$ $U(N)$ gauge theory with an adjoint and $N_f$ fundamental hypermultiplets. They obtained the free energy in the large $N$ limit at fixed $N/N_f$.}  The second method can be applied to many ${\cal N} \geq 2$ theories with M-theory duals \cite{Jafferis:2011zi,Martelli:2011qj,Cheon:2011vi,Gulotta:2011si,Gulotta:2011aa,Gulotta:2012yd}, but so far it can only be used to calculate $f_{3/2}$.  The third method has been generalized to certain ${\cal N} \geq 2$ supersymmetric theories with unitary gauge groups \cite{Marino:2012az};  in all these models, $Z_{S^3}$ is expressible in terms of an Airy function.  

We provide two extensions of the above methods.  We first extend method \eqref{second} to calculate the $k^{3/2}$ contribution to $f_{1/2}$ in \eqref{fABJM}, and provide a generalization to other SCFTs.  We then extend the Fermi gas approach \eqref{third} to SCFTs with orthogonal and symplectic gauge groups.  This method allows us to extract $f_{1/2}$ exactly for these theories, and we find agreement with results obtained using method \eqref{second}.  The extension of the Fermi gas approach to theories with symplectic and orthogonal gauge groups requires a fairly non-trivial generalization of the Cauchy determinant formula that we prove in the Appendix.  This formula allows us to write $Z_{S^3}$ as the partition function of non-interacting fermions that can move on half of the real line and obey either Dirichlet or Neumann boundary conditions at $x = 0$. We find that the result for $Z_{S^3}$ is again an Airy function.  

The rest of this paper is organized as follows.  In Section~\ref{sec:FieldTheories} we describe the field theories that we will consider in this paper.  These theories are not new.  They can be constructed in type IIA string theory using D2 and D6 branes, as well as O2 and O6 orientifold planes.  In Section~\ref{sec:LargeN} we extend the large $N$ expansion \eqref{second} to the next order.  In Section~\ref{sec:Fermi} we extend the Fermi gas approach \eqref{third} to our theories of interest.  We end with a discussion of our results in Section~\ref{sec:Discussion}. We include several appendices. In Appendix~\ref{MODULISPACE} we determine the moduli space of vacua using field theory techniques.  Appendix~\ref{app:FermiGas} provides a brief summary of the Fermi gas approach \cite{Marino:2011eh}. Appendix~\ref{app:Fourier} contain some details of our computations.  Lastly, in Appendix~\ref{app:Determinant} we prove the generalization of the Cauchy determinant formula used in the Fermi gas approach.  
\\

{\it Note added:} After this paper was published, it was pointed by~\cite{Assel:2015hsa}  that our treatment of SCFTs with orthogonal and symplectic gauge groups in the Fermi gas approach \eqref{third} missed a constant shift in the number of states below a certain energy.  This shift was calculated in~\cite{Assel:2015hsa,Okuyama:2015auc} using a more refined treatment of a Fermi gas on a half line than we offer here. We updated our results to incorporate this shift.

\section{Review of ${\cal N} = 4$ superconformal field theories and their string/M-theory description}\label{sec:FieldTheories}

\subsection{Brane construction and M-theory lift}
\label{BRANE}

We restrict ourselves to the simplest ${\cal N}=4$ superconformal field theories in $d=3$ with weakly-curved eleven-dimensional supergravity duals.  The field content of our theories of interest have an ${\cal N} = 4$ vectormultiplet with gauge group $U(N), \, O(2N),\, O(2N+1),$ or $USp(2N)$, a hypermultiplet transforming in a two-index tensor representation of the gauge group, and $N_f$ hypermultiplets transforming in the fundamental (vector) representation.  The two-index tensor representation can be the adjoint in the case of $U(N)$, or it can be a rank-two symmetric or anti-symmetric tensor representation in the other cases.

\begin{figure}[!h]
\begin{center}
\centering
        \begin{subfigure}[b]{0.48\textwidth}
                \centering
                \includegraphics[width=\textwidth]{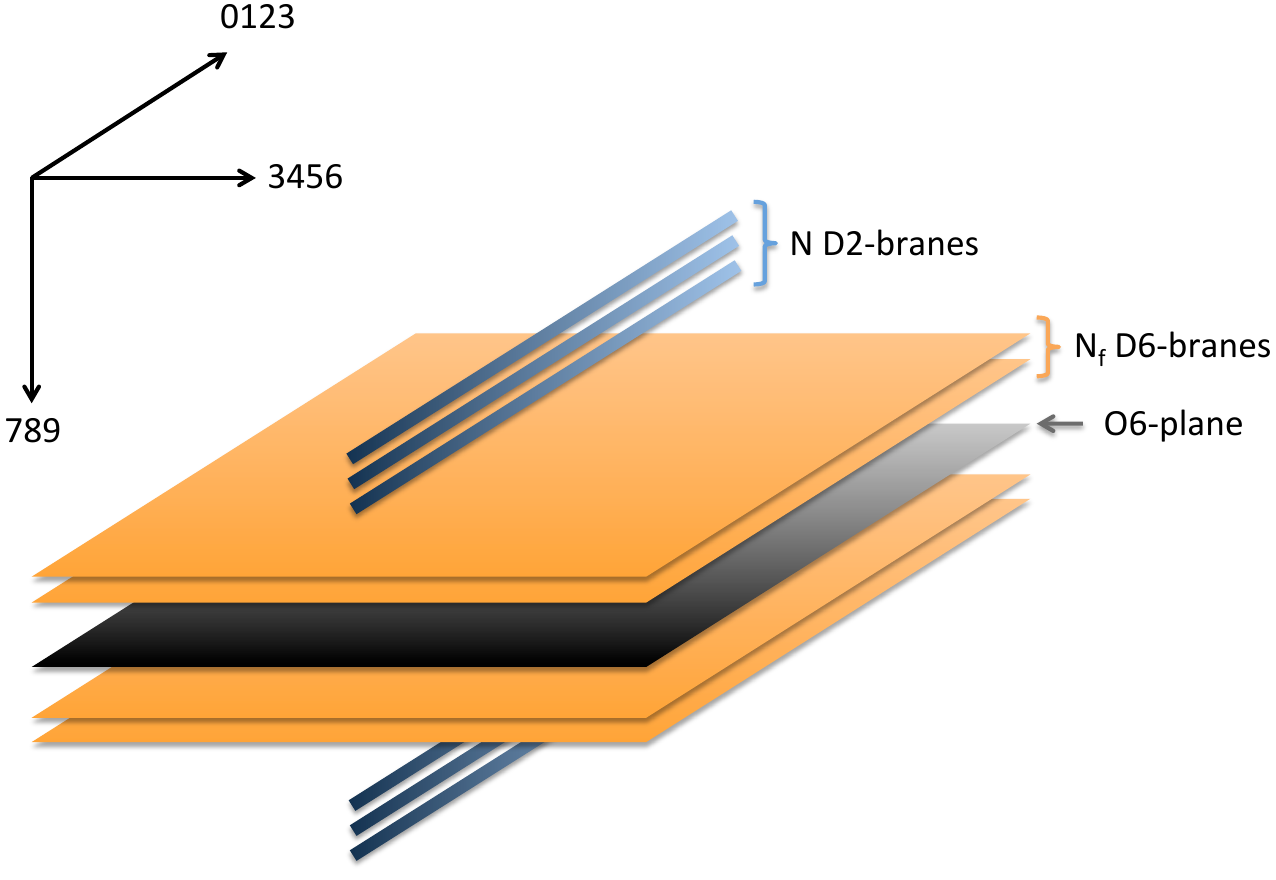}
                \caption{Brane construction with an O$6$-plane.}
                \label{fig:branes1}
        \end{subfigure}\hspace{0.5cm}
   \begin{subfigure}[b]{0.48\textwidth}
                \centering
                \includegraphics[width=\textwidth]{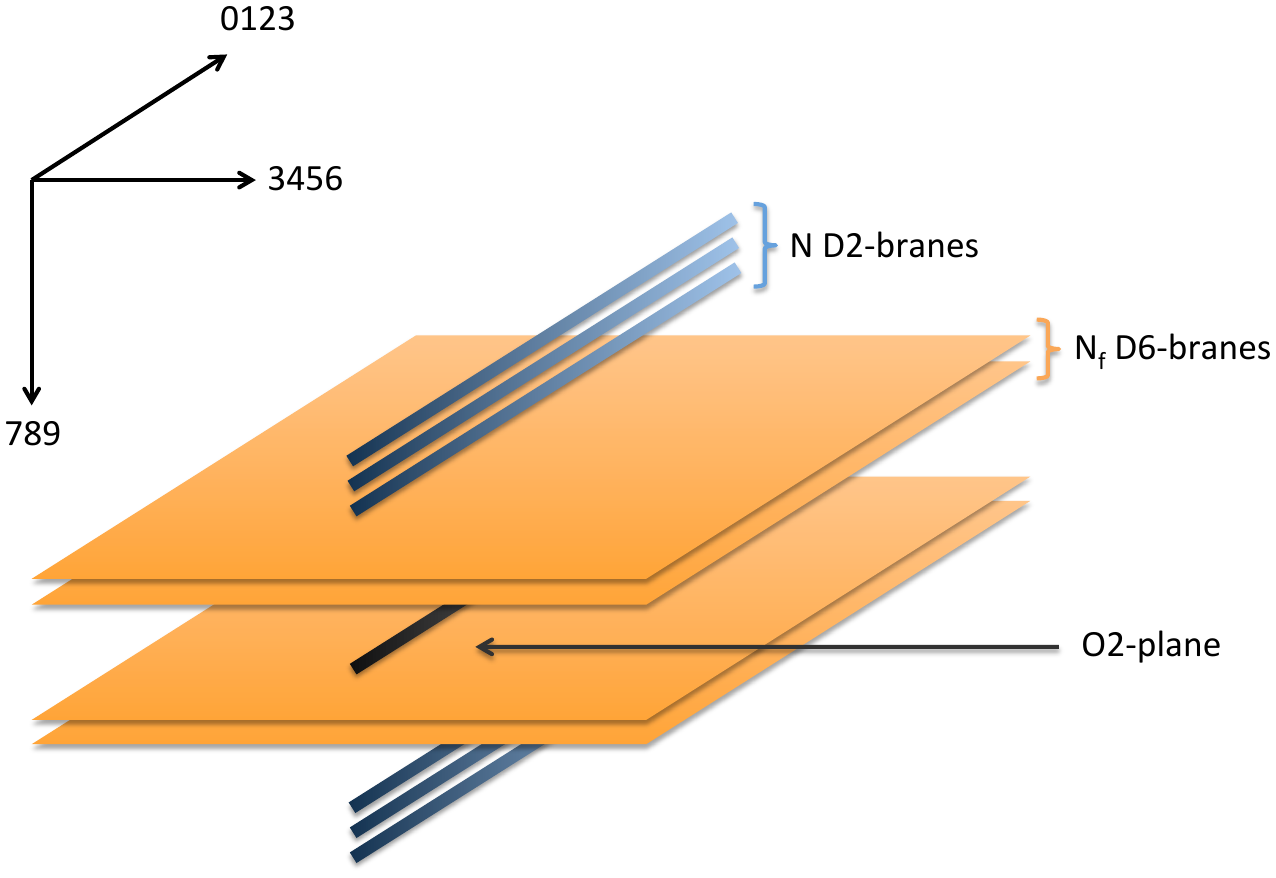}
                \caption{Brane construction with an O$2$-plane.}
                \label{fig:branes2}
        \end{subfigure}
                \caption{Type IIA brane construction of the theories considered. Exactly which figure applies, and what type of orientifold plane is needed can be read from Table~\ref{tab:Ingredients}.}
\label{fig:BraneConstruction}
\end{center}
\end{figure}

These SCFTs can be realized as low-energy effective theories on the intersection of various D-branes and orientifold planes in type IIA string theory as follows.  In all of our constructions, we consider D2-branes stretched in the $012$ directions, D6-branes stretched in the 0123456 directions, as well as O2-planes parallel to the D2-branes and O6-planes parallel to the D6-branes---See Table~\ref{tab:BraneDirections}.  Our constructions will have either an O2-plane or an O6-plane, but not both.  The gauge theory lives in the 012 directions, and the choice of gauge group and two-index tensor representation is dictated by the kind of O2 or O6-plane that is present.  The role of the D6-branes is to provide the fundamental hypermultiplet flavors.  See Figure~\ref{fig:BraneConstruction} for a picture of the brane configurations, and Table~\ref{tab:Ingredients} for which gauge theories correspond to which brane/orientifold constructions.
\begin{table}[!h]
\begin{center}
\begin{tabular}{c||cccccccccc}
Object & 0&1 &2&3&4&5&6&7&8&9\\
\hline
\hline
D2/O$2$ & $\bullet$ & $\bullet$& $\bullet$ &&&&&&&\\
\hline
D6/O$6$ & $\bullet$ & $\bullet$& $\bullet$ & $\bullet$ & $\bullet$& $\bullet$& $\bullet$&& &\\
\end{tabular}
\end{center}
\caption{The directions in which the ingredients extend are marked by $\bullet$.}\label{tab:BraneDirections}
\end{table}%

More precisely: 
\begin{itemize}
  \item $N$ D2-branes spanning the $012$ directions and $N_f$ D6-branes extending in the $0123456$ directions yields the ${\cal N} = 4$ $U(N)$ gauge theory with an adjoint hypermultiplet and $N_f$ fundamental hypermultiplets.
 \item To get the $O(2N)$ (or $O(2N+1)$) theory with an adjoint (antisymmetric tensor) hypermultiplet we add an O$2^-$-plane (or $\widetilde{\text{O}2}^-$-plane) coincident with the $2N$ D2-branes.\footnote{What we mean by this is that we have $N$ half D2-branes and their $N$ images. The $\widetilde{\text{O}2}^-$ can be thought of as having a half D2-brane stuck to an O$2^-$ plane, and hence naturally gives an $O(2N+1)$ gauge group.} The $2N_f$ D6-branes\footnote{In the case of orientifold planes, the D6-branes should be more correctly referred to as $N_f$ half D6-branes and their $N_f$ images under the orientifold action.} give $N_f$ fundamental flavors in the field theory living on the D2-branes.
 \item  If, on the other hand, we want to construct the $O(2N)$ (or $O(2N+1)$) theory with a symmetric tensor hypermultiplet we add an O$6^+$-plane coincident with the $2N_f$ D6-branes. To get the $O(2N)$ theory we need $2N$ D2-branes, while to get $O(2N+1)$ we need a half D2-brane to be stuck at the O$6^+$-plane.
 \item Similarly, to get the $USp(2N)$ theory with an adjoint (symmetric tensor) hypermultiplet we add an O$2^+$-plane coincident with $2N$ D2-branes.  The same theory can be obtained by using an  $\widetilde{\text{O}2}^+$-plane.\footnote{In a similar construction involving $2N$ D3-branes coincident with an O$3^+$ or with an $\widetilde{\text{O}3}^+$ plane one does obtain two distinct gauge theories with symplectic gauge groups denoted by $USp(2N)$ and $USp'(2N)$, respectively.  These theories differ in their spectra of dyonic line operators.} 
 \item To get the $USp(2N)$ gauge theory with an antisymmetric hypermultiplet, we should instead use an O$6^-$-plane.
 \item There are further ingredients in type IIA string theory, such as $\widetilde{\text{O}6}^{\pm}$-planes, but we do not use them in our constructions, because they do not yield 3-d SCFTs with known weakly-curved M-theory duals.\footnote{We do not consider $\widetilde{\text{O}6}^{\pm}$ planes in our brane constructions, as they require a non-zero cosmological constant~\cite{Hyakutake:2000mr,deBoer:2001px} in ten dimensions.  These orientifold planes therefore only exist in massive type IIA string theory and their M-theory lifts are unknown.   From the effective $2+1$-dimensional field theory perspective, an $\widetilde{\text{O}6}^{-}$-plane would introduce an extra fundamental half hypermultiplet compared to the O$6^-$ case.  The extra half hypermultiplet introduces a parity anomaly, which can be canceled by adding a bare Chern-Simons term.  This Chern-Simons term reduces the supersymmetry to ${\cal N}=3$~\cite{deBoer:2001px} and is related to the cosmological constant in ten dimensions.}\footnote{We remind the reader that it is impossible to have a half D2-brane stuck to an O6$^-$-plane, because the way the orientifold projection is implemented on the Chan-Paton factors requires an even number of such branes \cite{Gimon:1996rq}.}
\end{itemize}

\begin{table}[!h]
\begin{center}
\begin{tabular}{c||c|c|c|c|c|c|c||c}
$G\,+$ matter & D2 & D6 & O$2^-$ & $\widetilde{\text{O}2}^-$ & O$2^+$& O$6^-$& O$6^+$& Dual geometry AdS$_4\times X$\\
\hline
\hline
$U(N)\,+\text{adj}$ & $N$ & $N_f$ &&&&& &$S^7 / \Z_{N_f}$\\
\hline
$O(2N)\,+A$ & $2N$ & $2N_f$ &$\checkmark$&&&&&$(S^7/\hat{D}_{N_f})_\text{free}$\\
\hline
$O(2N)\,+S$  & $2N$ & $2N_f$ &&&&&$\checkmark$&$S^7/{\hat D}_{N_f+2}$\\
\hline
$O(2N+1)\,+A$  & $2N$ & $2N_f$ &&$\checkmark$&&&&$(S^7/\hat{D}_{N_f})_\text{free}$\\
\hline
$O(2N+1)\,+S$&  $2N+1$ & $2N_f$ &&&&&$\checkmark$&$S^7/{\hat D}_{N_f+2}$\\
\hline
$USp(2N)\,+A$& $2N$ & $2N_f$ &&&&$\checkmark$&&$S^7/{\hat D}_{N_f-2}$\\
\hline
$USp(2N)\,+S$ & $2N$ & $2N_f$ &&&$\checkmark$&&&$(S^7/\hat{D}_{N_f})_\text{free}$\\
\end{tabular}
\end{center}
\caption{The ingredients needed to construct a theory with gauge group $G$, $N_f$ fundamental flavors, and a two-index antisymmetric $(A)$ or symmetric $(S)$ hypermultiplet in Type IIA string theory. The dual M-theory background is also included.}\label{tab:Ingredients}
\end{table}%

The type IIA brane construction presented above can be straightforwardly lifted to M-theory, where one obtains $N$ M2-branes probing an $8$-(real)-dimensional hyperk\"ahler cone.\footnote{The M-theory description is valid at large $N$ and fixed $N_f$.  When $N_f$ is also large, a more useful description is in terms of type IIA string theory.} Indeed, if one ignores the D2-branes and orientifold planes for a moment, the configuration of $N_f$ separated D6-branes lifts to a configuration of $N_f$ unit mass Kaluza-Klein (KK) monopoles, and near every monopole core the spacetime is regular~\cite{Sen:1997zb}. $N_f$ coincident D6-branes correspond to coincident KK monopoles, whose core now has an $A_{N_f-1}$ singularity;  in other words, the transverse space to the monopole is $\C^2/\Z_{N_f}$ in this case. The infrared limit of the field theories living on the D2-branes is captured by M2-branes probing the region close to the core of the 11d KK monopole. Let us write the transverse directions to the M2-branes in complex coordinates. Let $z_1,z_2$ be the directions along which the KK monopole is extended, and $z_3,z_4$ be the directions transverse to it. Then the M2-branes probe the space $ \C^2\times\le(\C^2/\Z_{N_f}\ri)$~\cite{Aharony:2008ug}, where the $\Z_{N_f}$ action on the coordinates is given by
\es{ZNfAction}{
(z_3,z_4)\to e^{2\pi i\ov N_f}(z_3,z_4) \ .
}
The orbifold acts precisely in the direction of the M-theory circle, which therefore rotates $(z_3, z_4)$ by the same angle and is non-trivially fibered over the 7 directions transverse to the D2-branes.\footnote{Explicitly, the coordinates $x_3, \ldots, x_9$ transverse to the D2-branes can be identified with $(\Re z_1, \Im z_1, \Re z_2, \Im z_2, \Re (z_3 z_4^*), \Im (z_3 z_4^*), \abs{z_3}^2 - \abs{z_4}^2)$.  The M-theory circle is parameterized by $\psi = \frac 12 (\arg z_3 + \arg z_4) \in [0, 2 \pi)$, and \eqref{ZNfAction} identifies $\psi \sim \psi + 2\pi / N_f$. \label{MToTypeIIA}}

Back-reacting the $N$ M2-branes and taking the near horizon limit yields AdS$_4\times (S^7 / \Z_{N_f})$, where the $\Z_{N_f}$ action on $S^7$ is that induced from $\C^4$, namely \eqref{ZNfAction}. This orbifold action is not free, hence $S^7 / \Z_{N_f}$ is a singular space.  Since we have not included orientifold planes yet, this AdS$_4\times (S^7 / \Z_{N_f})$ background of M-theory is dual to the $U(N)$ theory with an adjoint and $N_f$ fundamental hypermultiplets. Note that for $N_f=1$ the monopole core is regular, the transverse space to the monopoles is $\C^2$, and the gravitational dual is M-theory on AdS$_4\times S^7$.  At low energies, M-theory on this background is dual to ABJM theory at Chern-Simons level $k=1$~\cite{Aharony:2008ug}; therefore, the $U(N)$ gauge theory with an adjoint and a flavor hypermultiplet described above is dual to ABJM theory at CS level $k=1$~\cite{Kapustin:2010xq}.

Introducing orientifolds in the type IIA construction corresponds to further orbifolding the 11d geometry.\footnote{We thank Oren Bergman and especially Ofer Aharony for helpful discussions on the lift of orientifolds to M-theory.}   The case of O$2$-planes is simpler:  the orbifold in 11d is generated by the action:
\es{O2Generator}{
\text{O2 lift:} &\qquad (z_1,z_2,z_3,z_4)\to (-z_1,-z_2,iz_4^*,-iz_3^*) \ .
} 
(See~\cite{Aharony:2008gk} for a similar orbifold action.)  This action can be derived from the fact that in type IIA an O$2$-plane acts both by flipping the sign of all the transverse coordinates as well as of the R-R one-form $A_1$.  This R-R one-form lifts to the off-diagonal components of the 11-d metric involving the M-theory circle and the type IIA coordinates (see for example \cite{Bergman:2001rp}), so in 11d the orientifold acts by a sign flip on the M-theory circle. Eq.~\eqref{O2Generator} then follows from the relations given in footnote~\ref{MToTypeIIA}.  We should combine the orbifold action \eqref{O2Generator} with \eqref{ZNfAction} (with $N_f\to2N_f$).  Together, the two generate the  dicyclic (binary dihedral) orbifold group, ${\hat D}_{N_f}$ of order $4N_f$.\footnote{Let us denote the O2 action in \eqref{O2Generator} by $a$ and the orbifold action \eqref{ZNfAction} (with $N_f \to 2 N_f$) by $b$. We then get the presentation of the dicyclic group ${\hat D}_{N_f}=\<a,b\vert\, b^{2N_f}=1,\, a^2=b^{N_f},\, ab=b^{2N_f-1}a\>$.} For $N_f=0$ there are no D6-branes, hence the orbifold group is just $\Z_2$. For $N_f=1$ the orbifold group is $\hat D_1 = \Z_4$.

In M-theory, we therefore have $N$ M2-branes probing a $\C^4 /  \hat D_{N_f}$ singularity, where $\hat D_{N_f}$ is generated by \eqref{ZNfAction} (with $N_f \to 2 N_f$) and \eqref{O2Generator}. In the near-horizon limit, the eleven dimensional geometry is AdS$_4\times (S^7/\hat{D}_{N_f})_\text{free}$.  The subscript ``free'' emphasizes that the orbifold action induced from \eqref{ZNfAction}--\eqref{O2Generator} on the $S^7$ base of $\C^4$ is free, and hence the corresponding eleven-dimensional background is smooth.  Note that the $\hat D_{N_f}$ orbifolds here are not the same as those in \cite{Aharony:2008gk} obtained from similar brane constructions.\footnote{The $N_f = 0$ case is special, because there are no D6-branes in this case.  In M-theory one obtains a pair of $\Z_2$ singularities corresponding to a pair of OM2 planes sitting at opposite points on the M-theory circle.  The gauge theory is simply ${\cal N} = 8$ SYM with $O(2N)$, $O(2N+1)$, or $USp(2N)$ gauge group, and just like ${\cal N} = 8$ SYM with gauge group $U(N)$, its infrared limit is non-standard.  We expect ${\cal N} = 8$ SYM with orthogonal or symplectic gauge group to flow to an ABJ(M) theory with Chern-Simons level $k=2$. \label{Nf0Footnote}}

The O$6$ case is more involved. The O$6^-$-plane lifts to Atiyah-Hitchin space in M-theory~\cite{Seiberg:1996bs,Seiberg:1996nz}. The O$6^-$-plane together with $2N_f$ coincident D6-branes away from the center of the Atiyah-Hitchin space can be thought of as a KK monopole with mass $(-4)$ (as the D6-brane charge of O$6^-$ is $(-4)$~\cite{Sen:1997kz}) and a KK monopole of mass $2N_f$, which we discussed above. When the D6-branes coincide with the O$6^-$-plane, we get a KK monopole of mass $2N_f-4$ (away from the center).  We should therefore consider the orbifold~\eqref{ZNfAction} with $N_f\to2N_f-4$.  In addition, the O$6$ plane yields an extra orbifold in 11d generated by
\es{O6Generator}{
 \text{O6 lift:} &\qquad (z_3,z_4)\to (iz_4^*,-iz_3^*) \ .
} 
As in the O2 case, this action can be derived from the fact that in type IIA an O$6$-plane acts by flipping the sign of all the transverse coordinates and of the R-R one-form $A_1$.  Together, \eqref{O6Generator} and \eqref{ZNfAction} (with $N_f \to 2 N_f - 4$) give a $D_{N_f}$ singularity. The corresponding orbifold group is again the dicyclic group, ${\hat D}_{N_f-2}$, so we have $N$ M2-branes probing a $\C^2 \times (\C^2 / \hat D_{N_f-2})$ transverse space.\footnote{The cases $N_f = 0, 1, 2$ are special.  When $N_f = 0, 1$, the 11-d geometry is smooth, and we therefore expect that the low-energy dynamics is the same as that of ABJM theory at level $k=1$.  When $N_f = 2$, the 11-d geometry has a pair of $\Z_2$ singularities.  Near each singularity the hyperk\"ahler space looks like $\C^2 \times (\C^2 / \Z_2)$. \label{Nf2Footnote}}

The M-theory lift of the O$6^+$ plane is a peculiar kind of $D_4$ singularity, perhaps with extra fluxes that prevent the possibility of blowing it up~\cite{Landsteiner:1997ei,Witten:1997bs}.  Further adding adding $2N_f$ D6-branes results in a $D_{N_f+4}$ singularity.  The corresponding orbifold group is ${\hat D}_{N_f+2}$, so in this case we have $N$ M2-branes probing a $\C^2 \times (\C^2 / \hat D_{N_f+2})$ transverse space.  Note that if we shift $N_f\to N_f+4$ in the O$6^-$ case, we get the same orbifold singularity as in the O$6^+$ case, perhaps with different torsion fluxes.   As we will see, the corresponding field theories do not have the same $S^3$ partition functions, so they are not dual to each other.

 For theories that are constructed with O$6$ planes, the near horizon limit of the M2-brane geometry is AdS$_4\times (S^7/{\hat D}_{N_f\pm2})$, where the $\hat D_{N_f \pm 2}$ action on $S^7$ is that induced from \eqref{ZNfAction} (with $N_f \to 2 N_f \pm 4$) and \eqref{O6Generator}. Within $\C^4$, the orbifold leaves the $\C^2$ at $z_3=z_4=0$ fixed, hence $S^7/{\hat D}_{N_f\pm2}$ is singular along the corresponding $S^3$.

In Appendix~\ref{MODULISPACE} we provide some evidence that the field theories mentioned above are indeed dual to M-theory on the backgrounds summarized in Table~\ref{tab:Ingredients} by computing the Coulomb branch of the moduli space.  In these moduli space computations an important role is played by certain BPS monopole operators that satisfy non-trivial chiral ring relations.  The Coulomb branch of the $U(N)$ theory with an adjoint and $N_f$ fundamental hypermultiplets is $( \C^2 \times (\C^2 / \Z_{N_f}) )^N / S_N$, where the symmetric group $S_N$ permutes the factors  in the product;  this branch of moduli space is precisely what is expected for $N$ M2-branes probing the hyperk\"ahler space $\C^2 \times (\C^2 / \Z_{N_f})$.  The Coulomb branch of the theories constructed from O2-planes is $(\C^4 / \hat D_{N_f} )^N / S_N$, again as expected for $N$ M2-branes probing $\C^4 / \hat D_{N_f} $.  The Coulomb branch of the theories constructed from O6-planes is $( \C^2 \times (\C^2 / \hat D_{N_f \pm 2}) )^N / S_N$ if the gauge group is $O(2N)$ or $USp(2N)$, matching the moduli space of $N$ M2-branes probing $\C^2 \times (\C^2 / \hat D_{N_f \pm 2})$.  If the gauge group is $O(2N+1)$ the moduli space has an extra factor of $\C^2$ corresponding to the half D2-brane stuck to the O$6^+$ plane that cannot move in the directions transverse to the orientifold plane.

It is worth pointing out that the moduli space computations in Appendix~\ref{MODULISPACE} provide agreement with the 11-d geometry only if certain details of the field theory are chosen appropriately.  For instance, the trace part of the symmetric tensor representations of $O(2N)$ and $O(2N+1)$ should be included, and so should the symplectic trace part of the anti-symmetric representation of $USp(2N)$.  In the $O(2N)$ cases, one finds agreement only if the gauge group is $O(2N)$, and not for $SO(2N)$---the two differ in a $\Z_2$ gauging of the global charge conjugation symmetry present in the $SO(2N)$ case.  In the $O(2N+1)$ case, the moduli space computation would yield the same answer as if the gauge group were $SO(2N+1)$.\footnote{The gauging of the charge conjugation symmetry in the $SO(2N+1)$ gauge theory does not seem to affect the dynamics provided that $2N+1 > N_f$.  When $2N+1 \leq N_f$, the $SO(2N+1)$ theory has baryonic operators of the form $q^{2N+1}$, where the color indices are contracted with the anti-symmetric tensor of $SO(2N+1)$.  These operators are odd under charge conjugation, and are therefore absent from the $O(2N+1)$ theory.  When $2N+1 > N_f$, however, the operator content of the $SO(2N+1)$ and $O(2N+1)$ gauge theories is the same. See also~\cite{Aharony:2013kma}.}
 
In all the cases, the eleven-dimensional geometry takes the form:
\es{Metric}{
ds^2&={R^2\ov 4}\, ds^2_\text{AdS$_4$}+R^2\, ds^2_X\ , \qquad R = \le(2^5 \pi^6  N\ov 3 \Vol(X)\ri)^{1/6} \ell_p\ ,\\
G_4&= \frac 38 R^3 \, \vol_\text{AdS$_4$} \ ,
}
where $R$ is the AdS radius, $\vol_\text{AdS$_4$}$ is the volume form on an AdS$_4$ of unit radius, $X$ is the internal seven-dimensional manifold (tri-Sasakian in this case), and $\ell_p$ is the Planck length.  This background should be accompanied by discrete torsion flux through a torsion three-cycle of $X$, but we do not attempt to determine this discrete torsion flux precisely.  Since the volume of $X$ is given by the volume of the unit $S^7$ divided by the order of the orbifold group, we predict using~\eqref{Gota} that
\es{Gota2}{
f_{3/2}={\sqrt{2}\pi\ov 3}\begin{cases}
N_f^{1/2} \qquad &\text{no orbifold,}\\
\le[4N_f\ri]^{1/2} \qquad & \text{O$2$},\\
\le[4(N_f\pm2)\ri]^{1/2} \quad & \text{O$6^\pm$} .
\end{cases}
}
These results will be reproduced by the field theory calculations presented in the remainder of this paper.  See Table~\ref{tab:CBValues}.

\subsection{Matrix model for the $S^3$ free energy}\label{sec:MatrixModel}

The $S^3$ partition function of $U(N)$ gauge theory with one adjoint and $N_f$ fundamental hypermultiplets can be written down using the rules summarized in \cite{Gulotta:2012yd}:
 \es{GotZUN}{
  Z &={1\ov 2^N \,N!} \int d^N x \, \frac{\prod_{i<j} \left[ 4 \sinh^2 \left(\pi (\lambda_i - \lambda_j) \right) \right] }
    {\prod_{i < j} \left[4 \cosh^2 \left(\pi (\lambda_i - \lambda_j) \right) \right]}\times \prod_i \frac{ 1}
{ \left(2 \cosh \left( \pi \lambda_i \right) \right)^{N_f}}\,.
 }
The normalization includes a division by the order of the Weyl group $|{\cal W}|=N!$ and the contributions from the $N$ zero weights in the adjoint representations.

The $S^3$ partition function for the theories with orthogonal and symplectic gauge groups is given by:
  \es{GotZ}{
  \widetilde Z &= {\cal C} \int d^N \lambda \, \frac{\prod_{i<j} \left[ 16 \sinh^2 \left(\pi (\lambda_i - \lambda_j) \right) \sinh^2 \left(\pi (\lambda_i + \lambda_j) \right) \right] }
    {\prod_{i < j} \left[16 \cosh^2 \left(\pi (\lambda_i - \lambda_j) \right) \cosh^2 \left(\pi (\lambda_i + \lambda_j) \right) \right]}\\
 &\times \prod_i \frac{ \left(4 \sinh^2 \left( \pi \lambda_i \right) \right)^a \left(4 \sinh^2 \left(2 \pi \lambda_i \right) \right)^b}
{ \left(4 \cosh^2 \left( \pi \lambda_i \right) \right)^{N_f+c} \left(4 \cosh^2 \left( 2\pi \lambda_i \right) \right)^{d}}\,.
 }
The constants $a$, $b$, $c$, $d$, and $ {\cal C}$ are given in Table~\ref{tab:Constants} for the various theories we study. The normalization ${\cal C}$ includes a division by the order of the Weyl group ${\cal W}$ (see Table~\ref{tab:Weyl}) and the contributions from in the zero weights the matter representations:
  \es{Normalization}{
    {\cal C} = \frac{1}{2^z \abs{\cal W}} \,,
  }
where $z$ is the total number of zero weights in the hypermultiplet representations.  In the $O(2N)$ and $O(2N+1)$ cases, \eqref{Normalization} should be multiplied by an extra factor of $1/2$ coming from the gauging of the $\Z_2$ charge conjugation symmetry that distinguishes the $O(2N)$ and $O(2N+1)$ gauge groups from $SO(2N)$ and $SO(2N+1)$, respectively.  In the rest of this paper, we find it convenient to rescale $\widetilde Z$ by a factor of $2^N$ and calculate instead
 \es{ZOSp}{
  Z = 2^N \widetilde Z \,.
 }

\begin{table}[!h]
\begin{center}
\begin{tabular}{c||c|c|c|c||c}
$G\,+$ matter & $a$ &$b$& $c$ & $d$ & $ {\cal C}$ \\
\hline
\hline
$O(2N)\,+A$ & 0 & 0 & 0 & 0 & $1/(2^{2N} N!)$ \\
\hline
$O(2N)\,+S$  & 0 & 0 &0 & 1 & $1/(2^{2N} N!)$  \\
\hline
$O(2N+1)\,+A$  & 1 & 0 & 1 & 0 & $1/(2^{2N+N_f + 1} N!)$ \\
\hline
$O(2N+1)\,+S$& 1 & 0 & 1 & 1 & $1/(2^{2N + N_f + 2} N!)$ \\
\hline
$USp(2N)\,+A$& 0 & 1 & 0 & 0 & $1/(2^{2N} N!)$ \\
\hline
$USp(2N)\,+S$ & 0 & 1 & 0 & 1 & $1/(2^{2N} N!)$ \\
\end{tabular}
\end{center}
\caption{The values of the constansts $a,b,c,$ and $d$ appearing in~\eqref{GotZ} for gauge group $G$, $N_f$ fundamental flavors, and a two-index antisymmetric $(A)$ or symmetric $(S)$ hypermultiplet.}\label{tab:Constants}
\end{table}%

\begin{table}[!h]
\begin{center}
\begin{tabular}{c||c}
$G$& $|{\cal W}|$\\
\hline
\hline
$U(N)$ & $ N!$\\
\hline
$SO(2N)$ & $2^{N-1}\, N!$\\
\hline
$SO(2N+1)$ & $2^{N}\, N!$\\
\hline
$USp(2N)$ & $2^{N}\, N!$\\
\end{tabular}
\end{center}
\caption{The order of the Weyl group,  $|{\cal W}|$, for various groups $G$.  In the case where the gauge group is $O(2N)$ or $O(2N+1)$, one should use the Weyl groups of $SO(2N)$ and $SO(2N+1)$ in \eqref{Normalization} and multiply the answer by an extra factor of $1/2$ coming from the gauging of the $\Z_2$ charge conjugation symmetry, as mentioned in the main text.}\label{tab:Weyl}
\end{table}%

The numerator in the integrand of~\eqref{GotZ} comes solely from the ${\cal N} = 4$ vectormultiplet;  note that an ${\cal N} = 4$ vector can be written as an ${\cal N} = 2$ vector and an ${\cal N} = 2$ chiral multiplet with R-charge $\Delta_\text{vec} = 1$, and only the ${\cal N} = 2$ vector gives a non-trivial contribution to the integrand.  The first factor in the denominator comes from the two-index hypermultiplet, while the additional factors come from both the two-index tensor and the $N_f$ fundamental hypermultiplets.  

Note that there is a redundancy in the parameters $a,b,$ and $c$. Using $\sinh 2 \lambda=2\sinh \lambda \, \cosh \lambda$, one can check that \eqref{GotZ} is invariant under
\es{Shifts}{
b\to b-\Delta\ ,  \qquad a\to a+\Delta\ , \qquad c\to c-\Delta \ ,
}
hence any expression involving $a$, $b$, and $c$ should only contain the combinations $c-a-2b$ or $a+b$. This requirement provides a nice check of our results.

\section{Large $N$ approximation}\label{sec:LargeN}

In this section we calculate the $S^3$ partition functions of the field theories presented above using the large $N$ approach of \cite{Herzog:2010hf}, which we extend to include one more order in the large $N$ expansion.  Explicitly, we do three computations.  In Section~\ref{sec:LargeNABJM} we present the computation for ABJM theory, whose $S^3$ partition function was given in \eqref{ZABJM}.  In Section~\ref{sec:LargeNUN}, we calculate the $F$-coefficient of the ${\cal N} = 4$ $U(N)$ gauge theory with one adjoint and $N_f$ fundamental hypermultiplets for which we wrote down the $S^3$ partition function in \eqref{GotZUN}.  Lastly, in Section~\ref{sec:LargeNUSp} we generalize this computation to theories with a symplectic or orthogonal gauge group, for which the $S^3$ partition function takes the form \eqref{GotZ} with various values of the parameters $a$, $b$, $c$, and $d$---see Table~\ref{tab:Constants}.

\subsection{ABJM theory}\label{sec:LargeNABJM}

At large $N$ one can calculate the $S^3$ partition function for ABJM theory \eqref{ZABJM} in a fairly elementary fashion using the saddle point approximation.  Let us write
 \es{ZSaddle}{
  Z = \frac{1}{(N!)^2} \int \prod_{i=1}^N \left( d \lambda_i d \tilde \lambda_i \right) \, e^{-F(\lambda_i, \tilde \lambda_j)} \,,
 }
for some function $F(\lambda_i, \tilde \lambda_j)$ that can be easily read off from \eqref{ZABJM}.  The factor of $(N!)^2$ that appears in \eqref{ZSaddle} is nothing but the order of the Weyl group ${\cal W}$, which in this case is $S_N \times S_N$, $S_N$ being the symmetric group on $N$ elements.  The saddle point equations are
 \es{Saddle}{
   \frac{\partial}{\partial \lambda_i} F(\lambda_i, \tilde \lambda_j) =  \frac{\partial}{\partial \tilde \lambda_j} F(\lambda_i, \tilde \lambda_j) = 0 \,.
 }
Since $F(\lambda_i, \tilde \lambda_j)$ is invariant under permuting the $\lambda_i$ or the $\tilde \lambda_j$ separately, the saddle point equations have a $S_N \times S_N$ symmetry.  For any solution of \eqref{Saddle} that is not invariant under this symmetry, as will be those we find below, there are $(N!)^2 - 1$ other solutions that can be obtained by permuting the $\lambda_i$ and the $\tilde \lambda_j$.  That our saddle point comes with multiplicity $(N!)^2$ means that we can approximate
 \es{ZApprox}{
  Z \approx e^{-F_*} \,,
 }
where $F_*$ equals the function $F(\lambda_i, \tilde \lambda_j)$ evaluated on any of the solutions of the saddle point equations.  In other words, the multiplicity of the saddle precisely cancels the $1/(N!)^2$ prefactor in \eqref{ZSaddle}.

The saddle point equations \eqref{Saddle} are invariant under interchanging $\tilde \lambda_i \leftrightarrow \lambda_i^*$, and therefore one expects to find saddles where $\tilde \lambda_i = \lambda_i^*$.   If one parameterizes the eigenvalues by their real part $x_i$, the density of the real part $\rho(x) =\frac{1}{N} \sum_{i=1}^N \delta(x - x_i)$ and $\lambda_i$ become continuous functions of $x$ in the limit $N \to \infty$.  The density $\rho(x)$ is constrained to be non-negative and to integrate to $1$.   Expanding $F(\lambda_i, \tilde \lambda_j)$ to leading order in $N$ (at fixed $N/k$), one obtains a continuum approximation:
\es{NExpABJM}{
\frac{F}{N^2}&= \int dx \ \rho(x) \int dx'\ \rho(x') \log\le|\cosh^2\le[\pi\le(\lam(x)-\widetilde\lam(x')\ri) \ri]\ov \sinh\Bigl[\pi\le(\lam(x)-\lam(x')\ri) \Bigr]\sinh\le[\pi\le(\widetilde\lam(x)-\widetilde\lam(x')\ri) \ri]\ri|\\
  &-i \frac{k}{N} \int dx \ \rho(x) \le(\lam(x)^2-\widetilde\lam(x)^2\ri)  + {\cal O}(1/N)   \,.
}
The corrections to this expression are suppressed by inverse powers of $N$.  In the $N\to\infty$ limit the saddle point approximation becomes exact, and to leading order in $N$ one can simply evaluate $F$ on the solution to the equations of motion following from \eqref{NExpABJM}.

At large $N/k$, one should further expand~\cite{Herzog:2010hf}:
\es{LambdaSaddle}{
\lam(x)=\sqrt{\frac Nk} \, x +i y(x) + \cdots\,,  \qquad \widetilde\lam(x)=\sqrt{\frac N k} \, x -i y(x) + \cdots \,,
}
with corrections suppressed by positive powers of $\sqrt{k/N}$.  Plugging \eqref{NExpABJM} into \eqref{NExpABJM} and expanding at large $N/k$, we obtain
\es{abjmlag}{
  \frac{F[\rho,y]}{N^2} &=\left(\frac kN\right)^{1/2} {\pi \ov 2} \int dx\ \le[ \rho^2\le(1-16y^2\ri)+ 8x \rho y \ri]\\
 &+ \left(\frac kN\right)^{3/2} \pi \int dx\ {1\ov 96}\rho\le(1-16y^2\ri)\le[64\rho' yy'+16\rho\le(3y'^2+2yy''\ri)-\le(1-16y^2\ri)\rho''\ri]  \\
 & +\dots  \ .
}
Note that the double integral in \eqref{NExpABJM} becomes a single integral in \eqref{abjmlag} after using the fact that, in the continuum limit \eqref{NExpABJM}, the scaling behavior \eqref{LambdaSaddle} implies that the interaction forces between the eigenvalues are short-ranged.  The expression in \eqref{abjmlag} should then be extremized order by order in $k/N$.  To leading order, the extremum was found in~\cite{Herzog:2010hf}:
\es{SPASolution}{
\rho(x)&=\begin{cases}
\sqrt{1\ov 2} \quad \text{for  $\abs{ x}\leq{1\ov\sqrt{2}}$}\,, \\
0 \qquad \text{otherwise} \,,
\end{cases}\\
y(x)&=\begin{cases}
\sqrt{1\ov 8}\, x \quad \text{for  $\abs{ x}\leq {1\ov\sqrt{2}}$} \,,\\
0 \qquad \text{otherwise} \,.
\end{cases}
}
This eigenvalue distribution only receives corrections from the next-to-leading term in the expansion \eqref{abjmlag}, so it is correct to plug \eqref{SPASolution} into \eqref{abjmlag} and obtain 
\es{ABJMFStar2}{
F_*= N^2 \left[ \left(\frac kN\right)^{1/2} {\sqrt2  \, \pi \ov 3} - \left(\frac kN\right)^{3/2}  {\pi \ov 24\sqrt{2}} + \ldots \right] +\dots \ .
}
If one wants to go to higher orders in the $k/N$ expansion, one would have to consider corrections to the eigenvalue distribution \eqref{SPASolution}. 

The result \eqref{ABJMFStar2} is in agreement with the Fermi gas approach \cite{Marino:2011eh}, when the latter is expanded at large $N/k$ and large $N$ as in \eqref{ABJMFStar2}.  The coefficients $f_{3/2}$ and $f_{1/2}$ of the $N^{3/2}$ and $N^{1/2}$ terms in the large $N$ expansion of the free energy obtained through the Fermi gas approach were given in~\eqref{fABJM}.  
Note that $F_*$ does not capture all the terms at ${\cal O}(N^{1/2})$, but only the contribution that scales as $k^{3/2}$.  This result is still meaningful, as the other terms of ${\cal O} (N^{1/2})$ in \eqref{fABJM}, coming from the fluctuations and finite $N$ corrections, have a different dependence on $k$.

\subsection{${\cal N}=4$ $U(N)$ gauge theory with adjoint and fundamental matter}\label{sec:LargeNUN}

We now move on to a more complicated example, namely the ${\cal N} = 4$ $U(N)$ gauge theory introduced in Section~\ref{sec:FieldTheories} whose $S^3$ partition function was given in \eqref{GotZUN}.  Let us denote
 \es{ZToF}{
  Z = \frac{1}{\abs{{\cal W}}}  \int d^N \lambda\, e^{-F(\lambda_i)} \,.
 }
Explicitly, we have
 \es{FExplicit}{
  F(\lambda_i) =  - \sum_{i<j} \log \tanh^2 (\pi (\lambda_i - \lambda_j))  - \sum_i \log \frac{1 }{\left(2 \cosh (\pi \lambda_i) \right)^{N_f}} \,.
 }
As in the ABJM case, every saddle comes with a degeneracy equal to the order of the Weyl group ($S_N$ in this case), so we can approximate $Z \approx e^{-F_*}$, where $F_*$ equals $F(\lambda_i)$ evaluated on any given solution of the saddle point equations $\partial F / \partial \lambda_i = 0$. 

In the $U(N)$ gauge theory the eigenvalues are real, and in the $N\to\infty$ limit we again introduce a density of eigenvalues $\rho(x)$. We will be interested in taking $N$ to infinity while working in the Veneziano limit where $t \equiv N/N_f$ is held fixed and then taking the limit of large $t$.  At large $N$, the free energy is a functional of $\rho(x)$:
 \es{FApproxUN}{
  \frac{F[\rho]}{N^2} &= \int dx\, \rho(x) \int dx'\, \rho(x') \log 
  \abs{\coth \left(\pi (\lambda(x) - \lambda(x') ) \right) } \\
   &+ \frac{1}{t} \int dx\, \rho(x)  \,
      \log \left( 2\cosh (\pi \lambda(x)) \right) \,.
 }

As in the ABJM case, the appropriate scaling at large $t$ is $\lambda \propto \sqrt{t}$, so we can define
 \es{Larget}{
  \lambda(x) = \sqrt{t}\, x \,.
 }
It is convenient to further introduce another parameter $T$ and write \eqref{FApproxUN} as
 \es{FApproxTUN}{
  \frac{F[\rho]}{N^2} &= \int dx\, \rho(x) \int dx'\, \rho(x') \log 
  \abs{\coth \left(\pi \sqrt{t} (x - x') \right) } \\
   &+ \frac{1}{\sqrt{t}} \int dx\, \frac{\rho(x)}{\sqrt{T}}  \,
      \log \left( 2\cosh (\pi \sqrt{T} x) \right) 
      \,.
 }
Of course, we are eventually interested in setting $T = t$, but it will turn out to be convenient to have two different parameters and expand both at large $t$ and large $T$.  Expanding in $t$ we get
 \es{FApproxTExpandUN}{
  \frac{F[\rho]}{N^2} &= \frac{\pi}{4} \frac{1}{\sqrt{t}} \int dx\, \rho(x)^2 - \frac{\pi}{192} \frac{1}{t^{3/2}} \int dx\, \rho'(x)^2  + o(t^{-3/2}) \\
   &+ \frac{1}{\sqrt{t}} \int dx\, \frac{\rho(x)}{\sqrt{T}}  
     \, \log \left( 2\cosh (\pi \sqrt{T} x) \right)  \,.
 }

If we assume that $\rho$ is supported on $[-x_*, x_*]$ for some $x_*>0$, we should extremize \eqref{FApproxTExpandUN} order by order in $N$ under the condition that $\rho(x) \geq 0$ and that
 \es{rhoNorm}{
  \int_{-x_*}^{x_*} dx\, \rho(x) = 1 \,.
 }
We can impose the latter condition with a Lagrange multiplier and extremize
 \es{tildeF}{
  \frac{\widetilde F[\rho]}{N^2} = \frac{F[\rho]}{N^2} - \pi \frac{\mu}{\sqrt{t}} \left(\int dx\, \rho(x) - 1 \right) 
 } 
instead of \eqref{FApproxTExpandUN}.

\subsubsection{Leading order result}

To obtain the leading order free energy we can simply take the limit $T\to \infty$ in \eqref{FApproxTExpandUN} and ignore the $1/t^{3/2}$ term in the first line of \eqref{FApproxTExpandUN}. The free energy takes the form
 \es{FApproxTExpandUN2}{
  \frac{F[\rho]}{N^2} &=  \frac{1}{\sqrt{t}}\int dx\, \le[\frac{\pi}{4}\,\rho(x)^2 + \pi x\,\rho(x) \ri] \,.
 }
The normalized $\rho(x)$ that minimizes \eqref{FApproxTExpandUN2} is
 \es{GotrhoSymmUN}{
  \rho(x) = \begin{cases}
   (x_* - \abs{x}) / x_*^2 &  \abs{x} \leq x_* \,, \\
   0 & \text{otherwise,} 
  \end{cases}  \qquad
   x_* \equiv \frac{1}{\sqrt{2 }} \,.
 } 

The value of $F$ we obtain from \eqref{GotrhoSymmUN} is
 \es{GotFUN}{
  \frac{F _*}{N^2}= \frac{\pi \sqrt{2}}{3} \frac{1}{\sqrt{t}} + o(t^{-1/2}) \,.
 }
After writing $t = N / N_f$, one can check that this term reproduces the expected $N^{3/2}$ behavior of a SCFT dual to AdS$_4 \times S^7 / \Z_{N_f}$.

\subsubsection{Subleading corrections}

To obtain the  $t^{-3/2}$ term in \eqref{GotFUN} we should find the $1/T$ corrections to the extremum of the $t^{-1/2}$ terms in \eqref{FApproxTExpandUN}, and we should evaluate the $t^{-3/2}$ term in \eqref{FApproxTExpandUN} by plugging in the leading result \eqref{GotrhoSymmUN}.

Focusing on the $t^{-1/2}$ terms first, the equation of motion for $\rho$ gives
 \es{LeadingTUN}{
  \rho(x) = 2\mu - \frac{2}{\pi \sqrt{T}}  
      \log \left( 2\cosh (\pi \sqrt{T} x) \right) \,.
 }
Up to exponentially small corrections (at large $T$), the normalization condition \eqref{rhoNorm} fixes $\mu$ to
 \es{muTUN}{
  \mu =  {x_*\ov 2} + \frac{1}{x_*} \left(\frac 14 + \frac{1}{24 T} \right) \,.
 }
 Plugging this expression back into $F[\rho]$ and minimizing with respect to $x_*$, one obtains
  \es{GotxsUN}{
  x_* = \sqrt{{1\ov 2} +{1  \ov 12 T}} \,,
 }
again only up to exponentially suppressed corrections. 

Then \eqref{FApproxTExpandUN} evaluates to 
 \es{FFinalUN}{
  \frac{F_*}{N^2} = \frac{ \pi \sqrt{2}}{3} \frac{1}{\sqrt{t}} + \frac{ \pi  }{6\sqrt2} \frac{1}{T\sqrt{t}}
   - \frac{\pi }{24\sqrt2 } \frac{1}{t^{3/2}} + \ldots \,,
 }
 where we included the $t^{-3/2}$ term.  We see now that if we had taken $T\to \infty$ directly in \eqref{FApproxTExpandUN} we would have missed the second term in \eqref{FFinalUN}.  Setting $T = t = N / N_f$, we obtain
 \es{FFinal2UN}{
   F_*= \frac{ \pi \sqrt{2N_f}}{3} N^{3/2} + \frac{\pi N_f ^{3/2}}{8\sqrt2} N^{1/2} + \dots \,.
 }
In analogy with the ABJM case we expect that fluctuations and finite $N$ corrections will contribute to the free energy starting at $N^{1/2}$ order. However, they will have different $N_f$ dependence then the term~\eqref{FFinal2UN}, and the saddle point computation can be thought of as the first term in the large $N_f$ expansion.\footnote{One could think of $t = N/N_f$ as the analog of the 't Hooft coupling in this case. }  These expectations will be verified in the Fermi gas approach in Section~\ref{sec:Fermi}.

\subsection{${\cal N}=4$ gauge theories with orthogonal and symplectic gauge groups}\label{sec:LargeNUSp}

As a final example, let us discuss the ${\cal N} = 4$ theories with symplectic and orthogonal gauge groups for which the $S^3$ partition function was written down in \eqref{GotZ}.  (See Table~\ref{tab:Constants} for the values of the constants $a$, $b$, $c$, $d$, and ${\cal C}$.)  In this case one can define $F(\lambda_i)$ just as in \eqref{ZToF}.  The saddle point equations $\partial F / \partial \lambda_i$ are now invariant both under permutations of the $\lambda_i$ and under flipping the sign of any number of $\lambda_i$.   In particular, from any solution of the saddle point equations one can construct other solutions by flipping the sign of any number of $\lambda_i$.  We can therefore restrict ourselves to saddles for which $\lambda_i \geq 0$ for all $i$.  If $F_*$ is the free energy of any such saddle, we have $Z \approx  e^{-F_*}$, up to a ${\cal O}(N^0)$ normalization factor coming from the constant ${\cal C}$ in \eqref{GotZ} that we will henceforth ignore.

Instead of extremizing $F(\lambda_i)$ with respect to the $N$ variables $\lambda_i$, $i = 1, \ldots, N$, it is convenient to introduce $2N$ variables $\mu_i$, $i = 1, \ldots, 2N$, and extremize instead
 \es{FInstead}{
  F(\mu_i) =-\frac 12 \sum_{i<j} \log \tanh^2 (\pi (\mu_i - \mu_j))  - \sum_i \log \abs{\frac{\left(2 \sinh (\pi \mu_i) \right)^a \left(2 \sinh (2\pi \mu_i) \right)^{b-1}  }{\left(2 \cosh (\pi \mu_i) \right)^{N_f + c} \left(2 \cosh (2\pi \mu_i) \right)^{d-1} } } 
 }
under the constraint $\mu_{i + N} = - \mu_i$.  In the case at hand, one can actually drop this constraint, because the extrema of the unconstrained minimization of $F(\mu_i)$ satisfy $\mu_{i + N} = - \mu_i$ (after a potential relabeling of the $\mu_i$).

If the $\mu_i$ are large, then extremizing \eqref{FInstead} is equivalent up to exponentially small corrections to extremizing 
 \es{FInstead2}{
  F(\mu_i) = \frac 12 \left[ - \sum_{i<j} \log \tanh^2 (\pi (\mu_i - \mu_j))  - \sum_i \log \frac{1 }{\left(2 \cosh (\pi \mu_i) \right)^{2 \K}} \right] \,,
 }
where
 \es{NftDef}{
  \K&\equiv N_f +c+2d-a-2b \,.
 }
We performed a similar extremization  problem in the previous section.  From comparing \eqref{FInstead2} with \eqref{FExplicit}, we see that the extremum of \eqref{FInstead2} can be obtained after replacing $N_f \to 2 \K$, $N \to 2N$ in \eqref{FFinal2UN} and multiplying the answer by $1/2$:
\es{FFinal3}{
F_*= \frac{2 \pi \sqrt{2\K}}{3} N^{3/2} + \frac{\pi \K^{3/2}}{4 \sqrt{2}} N^{1/2} +\dots \ .
}
The first term reproduces the expected $N^{3/2}$ behavior of an SCFT dual to AdS$_4 \times X$ where $X$ is an orbifold of $S^7$ of order $4\K$, in agreement with \eqref{Gota2}.  We will reproduce \eqref{FFinal3} from the Fermi gas approach in the following section, where we will also be able to calculate the other terms of order $N^{1/2}$ that have a different $\K$ dependence from the one in \eqref{FFinal3}.

\section{Fermi gas approach}\label{sec:Fermi}

\subsection{${\cal N}=4$ $U(N)$ gauge theory with adjoint and fundamental matter}

For SCFTs with unitary gauge groups and ${\cal N} \geq 3$ supersymmetry, the Fermi gas approach of \cite{Marino:2011eh,Marino:2012az} relies on the determinant formula
 \es{CauchyCoshMain}{
  \frac{\prod_{i<j} \left[ 4 \sinh \left(\pi (x_i - x_j)  \right) \sinh \left(\pi (y_i - y_j)  \right) \right] }
   {\prod_{i, j} \left[2 \cosh\left(\pi (x_i - y_j \right) \right]}
    = \det \frac{1}{2 \cosh \left( \pi (x_i - y_j) \right)} \,,
 } 
which is nothing but a slight rewriting of the Cauchy determinant formula
 \es{CauchyMain}{
  \frac{\prod_{i<j} (u_i - u_j) (v_i - v_j) }{\prod_{i, j} (u_i + v_j)} = \det \frac{1}{u_i + v_j} \,.
 }
that holds for any $u_i$ and $v_i$, with $i = 1, \ldots, N$.  Eq.~\eqref{CauchyCoshMain} can be obtained from \eqref{CauchyMain} by writing $u_i = e^{2 \pi x_i}$ and $v_i = e^{2 \pi y_i}$.

Using~\eqref{CauchyCoshMain} in the particular case $y_i = x_i$, we can write~\eqref{GotZUN} in the form
 \es{ZDetUN}{
  Z = \frac{1}{N!} \int d^N x \, \prod_i  \frac{1}{\left( 4 \cosh^2 (\pi x_i) \right)^{N_f}}
   \det \frac{1}{2 \cosh \left( \pi (x_i - x_j) \right) } \,.
 } 
$Z$ can then be rewritten as the partition function of an ideal Fermi gas of $N$ noninteracting particles, namely
 \es{ZDensity}{
  Z = \frac{1}{N!} \sum_{\sigma \in S_N} (-1)^\sigma \int d^N x\, \prod_i \rho\left(x_i, x_{\sigma(i)} \right) \,,
 } 
where $\rho(x_1, x_2) \equiv \langle x_1 | \hat \rho | x_2 \rangle$ is the one particle density matrix, and the sum is over the elements of the permutation group $S_N$. We can read off the density matrix by comparing \eqref{ZDensity} with \eqref{ZDetUN}.  In the position representation, $\rho$ is given by 
 \es{DenistyMatrixUN}{
  \rho(x_1, x_2) = \frac{1}{\left( 2 \cosh (\pi x_1) \right)^{N_f/2}} \,
    \frac{1}{\left( 2 \cosh (\pi x_2) \right)^{N_f/2}} 
    \times  \frac{1}{2 \cosh \left( \pi (x_1 - x_2) \right)} \,.
 }
We can put this expression into a more useful form by writing it more abstractly in terms of the position and momentum operators, $\hat x$ and $\hat p$, as
 \es{rhohatUN}{
  \hat \rho = e^{- U(\hat x)/2} e^{- T(\hat p)} e^{-U(\hat x)/2} \,,
 }
In units where $h=1$, which imply $[\hat x, \hat p]= i / (2 \pi)$, one can show as in~\eqref{MomRepTerm} that
 \es{GotUTUN}{
   U(x) = \log \left( 2 \cosh (\pi  x) \right)^{N_f} \,, \qquad
   T(p) = \log \left(2 \cosh (\pi  p) \right) \,.
 }
We then rescale $x\equiv y / (2\pi N_f)$ and $p \equiv k / (2 \pi)$ to get
 \es{rhohat2UN}{
    \hat \rho={1\ov 2\pi N_f} e^{- U(\hat y)/2} e^{-T(\hat k)} e^{- U(\hat y)/2} \,,
 }
where
 \es{GotUT2UN}{
   U(y) = \log \left( 2 \cosh \le({y\ov 2 N_f} \ri) \right)^{N_f} \,, \qquad
    T(k) = \log \left(2 \cosh \le(k\ov 2\ri) \right) \,.
 } 
 
The rescaling was motivated by the following nice properties:
\es{comasymUN}{
 \le[\hat y, \hat k\ri]&= 2\pi i N_f \ , \\
 U(y) &\to {y\ov 2} \qquad (y\to \infty)\ ,\\
T(k) &\to {k\ov 2} \qquad (k\to \infty) \ .
 \qquad 
}
We identify $\hbar=2\pi N_f$, and perform a semiclassical computation of the canonical free energy of the Fermi gas. In Appendix~\ref{app:FermiGas} we give a brief review of the relevant results from~\cite{Marino:2011eh}.  These results enable us to calculate the free energy from the above ingredients. In summary, we calculate the Fermi surface area as a function of the energy for the Wigner Hamiltonian~\eqref{WignerHamiltonian}. In the semiclassical approximation, to zeroth order the phase space volume  enclosed by the Fermi surface is:
\es{phsp0}{
V_0&=8 E^2 \ ,
}
while the corrections are:
\es{phsp1}{
\Delta V&=4\le[\int_0^\infty dy \, \le(y-2U(y)\ri)+\int_0^\infty dp \, \le(k-2T(k)\ri)+{\hbar^2\ov 24}\int_0^\infty dy \, U''(y)-{\hbar^2\ov 48}\int_0^\infty dk \, T''(k)\ri] \ .
}
We can perform the calculation and conclude that $n(E)$ defined in~\eqref{NumberEigen} takes the form:
\es{phpstotUN}{
n(E)&={V\ov 2\pi \hbar}={V_0+\Delta V\ov 2\pi \hbar}={2E^2 \ov \pi^2 N_f}-{ N_f \ov 8}-{1\ov 6N_f} \ .    
}
In~\eqref{nAsymp} we parametrized the $E$ dependence of $n(E)$ as
\es{nAsympMain}{
n(E)&=C\,E^2+n_0+{\cal O}\le(E\, e^{-E}\ri) \ , \\
B&\equiv n_0+{\pi^2 C\ov 3}\ , 
}
so from~\eqref{phpstotUN} we can read off
\es{CBUN}{
C&={2 \ov \pi^2 N_f} \ ,\qquad  B=-{ N_f \ov 8}+{1\ov 2N_f}\ , 
}  
and the partition function takes the form~\cite{Marino:2011eh}
\es{JAsympMain}{
Z(N)&={\cal A}(N_f) \, {\rm Ai}\le[C^{-1/3}(N-B)\ri]+ {\cal O}\le(e^{-\sqrt{N}}\ri)\ .
}
${\cal A}(N_f)$ is an $N$-independent constant that our approach only determines perturbatively for small $N_f$, and we are not interested in its value. Expanding the $F=-\log Z$ we obtain:
\es{FExpand}{
F&= f_{3/2} N^{3/2} + f_{1/2} N^{1/2} + \ldots\ , \qquad f_{3/2}={2\ov 3 \sqrt{C}}\ ,\qquad f_{1/2}=-{B\ov \sqrt{C}} \ .
}
We conclude that the free energy goes as:
\es{FUN}{
F&= \frac{ \pi \sqrt{2N_f}}{3} N^{3/2} + \frac{\pi }{\sqrt2}\le({N_f ^{3/2}\ov 8}-{1\ov 2 \sqrt{N_f}}\ri) N^{1/2} +\dots  \ .
}
The Fermi gas computation is in principle only valid in the semiclassical, small $\hbar$, i.e.~small $N_f$ regime.  However, because the small $N_f$ series expansions terminate, we obtain the exact answer. Then we can compare to the matrix model result~\eqref{FFinal2UN} valid at large $N_f$, and find perfect agreement to leading order in $N_f$.\footnote{Grassi and Mari\~no informed us that they calculated the free energy of this theory in the large $N$, fixed $N/N_f$ limit using method \eqref{first} discussed in Section~\ref{sec:Intro}.  Their result is $$F=\frac{ \pi \sqrt{2N_f}}{3} N^{3/2}\le(1+{N_f\ov 8 N}\ri)^{3/2}$$ up to exponentially small corrections in $N/N_f$ and subleading terms in $1/N$.  This expression agrees with the large $N$, fixed $N/N_f$ limit of the Fermi gas result~\eqref{CBUN}--\eqref{JAsympMain} of this section. We thank Marcos Mari\~no for sharing these results with us. }

As discussed in Section~\ref{sec:FieldTheories}, at $N_f=1$ the $U(N)$ theory is dual to ABJM theory at $k=1$, and the free energy computation in both representations should give the same result~\cite{Kapustin:2010xq}. Plugging $k=1$ into~\eqref{FExpansion}  and~\eqref{fABJM} indeed gives~\eqref{FUN} with $N_f=1$.

\subsection{${\cal N}=4$ gauge theories with orthogonal and symplectic gauge groups}

To generalize the Fermi gas approach to SCFTs with orthogonal and symplectic gauge groups, one needs the following generalization of the Cauchy determinant formula \eqref{CauchyMain}:
 \es{CauchyGenMain}{
  \frac{\prod_{i<j} (u_i - u_j) (v_i - v_j) (u_i u_j - 1) (v_i v_j - 1)}{\prod_{i, j} (u_i + v_j) (u_i v_j + 1)} = \det \frac{1}{(u_i + v_j)(u_i v_j + 1)} \,,
 }
which holds for any $u_i$ and $v_i$, with $i = 1, \ldots, N$.\footnote{After completing this paper, it was pointed out to us by Miguel Tierz that this determinant formula can be found in the literature.  See, for example, \cite{2000math......8184K}.}  Upon writing $u_i = e^{2 \pi x_i}$ and $v_i = e^{2 \pi y_i}$, this expression becomes
 \es{CauchyGenCoshMain}{
   \frac{\prod_{i<j} \left[ 16 \sinh \left(\pi (x_i - x_j)  \right) \sinh \left(\pi (y_i - y_j)  \right) 
     \sinh \left(\pi (x_i + x_j)  \right) \sinh \left(\pi (y_i + y_j)  \right) \right] }
   {\prod_{i, j} \left[4 \cosh\left(\pi (x_i - y_j \right) \cosh\left(\pi (x_i + y_j \right) \right]} \\
    = \det \frac{1}{4 \cosh \left( \pi (x_i - y_j) \right) \cosh \left( \pi (x_i + y_j) \right)} \,.
 } 
In addition to this generalization of the Cauchy determinant formula, our analysis involves an extra ingredient.  The one-particle density matrix of the resulting Fermi gas will be expressible not only just in terms of the usual position and momentum operators $\hat x$ and $\hat p$ as before, but also in terms of a reflection operator $R$ that we will need in order to project onto symmetric or anti-symmetric wave-functions on the real line.

Using~\eqref{CauchyGenCoshMain} in the particular case $y_i = x_i$, one can rewrite~\eqref{GotZ} as
\es{ZDet}{
  Z = 2^N {\cal C} \int d^N x \, \prod_i  \frac{ \le(4\sinh^2 (\pi x_i)\ri)^{a}\le(4\sinh^2 (2 \pi x_i)\ri)^{b}}{\left( 4 \cosh^2 (\pi x_i) \right)^{N_f+c}\, \left( 4 \cosh^2 (2\pi x_i) \right)^{d-1/2}} \\
\times 
   \det \frac{1}{4 \cosh \left( \pi (x_i - x_j) \right) \cosh \left( \pi (x_i + x_j) \right)} \,.
 } 
As in~\eqref{ZDensity} we recognize the appearence of the partition function of an ideal Fermi gas of $N$ noninteracting particles, and can read off the one-particle density matrix $\hat \rho$ from comparing \eqref{ZDensity} with \eqref{ZDet}.  From Table~\ref{tab:Weyl} we see that $2^{N}\, {\cal C} \approx 1/(2^N N!)$, up to a ${\cal O}(N^0)$ pre-factor that we will henceforth ignore.   In the position representation, $\rho(x_1, x_2) \equiv \langle x_1 | \hat \rho | x_2 \rangle$ is given by 
  \es{DenistyMatrix}{
  \rho(x_1, x_2) &= \frac12 \sqrt{\frac{ \le(4\sinh^2 (\pi x_1)\ri)^{a}\le(4\sinh^2 (2 \pi x_1)\ri)^{b}}{\left( 4 \cosh^2 (\pi x_1) \right)^{N_f+c}\, \left( 4 \cosh^2 (2\pi x_1) \right)^{d-1/2}}} 
    \sqrt{\frac{ \le(4\sinh^2 (\pi x_2)\ri)^{a}\le(4\sinh^2 (2 \pi x_2)\ri)^{b}}{\left( 4 \cosh^2 (\pi x_2) \right)^{N_f+c}\, \left( 4 \cosh^2 (2\pi x_2) \right)^{d-1/2}}} \\
    &\times  \frac{1}{4 \cosh \left( \pi (x_1 - x_2) \right) \cosh \left( \pi (x_1 + x_2) \right)} \,.
 }

To put this expression in a more useful form, we note that if we set $h = 1$, we can write
 \es{MomRepCosh}{
   \frac{4 \cosh (\pi x_1) \cosh(\pi x_2) }{4 \cosh \left( \pi (x_1 - x_2) \right) \cosh \left( \pi (x_1 + x_2) \right)}
   = \left\langle x_1 \Bigg\vert \frac{1 + R}{2 \cosh (\pi \hat p)}  \Bigg\vert  x_2 \right\rangle \,,
 }
where $R$ is the reflection operator that sends $x \to -x$.  For the derivation of this identity see Appendix~\ref{app:Fourier}. Then we can write
 \es{rhohat}{
  \hat \rho = e^{- U_+(\hat x)/2} e^{-T(\hat p)} \le({1+R\ov 2}\ri) e^{- U_+(\hat x)/2} \,,
 }
where
  \es{GotUT}{
   U_+(x) = \log \frac{\left( 4 \cosh^2 (\pi x) \right)^{N_f+c+1}\, \left( 4 \cosh^2 (2\pi x) \right)^{d-1/2}}{ \le(4\sinh^2 (\pi x)\ri)^{a}\le(4\sinh^2 (2 \pi x)\ri)^{b}} \,, \qquad
   T(p) = \log \left(2 \cosh (\pi  p) \right) \,.
 }

Similarly, we could use the identity
 \es{MomRepSinh}{
  \frac{4 \sinh (\pi x_1) \sinh(\pi x_2) }{4 \cosh \left( \pi (x_1 - x_2) \right) \cosh \left( \pi (x_1 + x_2) \right)}
   = \left\langle x_1 \Bigg\vert \frac{1 - R}{2 \cosh (\pi \hat p)}  \Bigg\vert  x_2 \right\rangle \ ,
 }
and write
 \es{rhohat2}{
  \hat \rho = e^{- U_-(\hat x)/2} e^{-T(\hat p)} \le({1-R\ov 2}\ri)e^{- U_-(\hat x)/2} \,,
 }
with
  \es{GotUT2}{
  U_-(x) = \log \frac{\left( 4 \cosh^2 (\pi x) \right)^{N_f+c}\, \left( 4 \cosh^2 (2\pi x) \right)^{d-1/2}}{ \le(4\sinh^2 (\pi x)\ri)^{a-1}\le(4\sinh^2 (2 \pi x)\ri)^{b}}  \,,
 }
and the same expression for $T(p)$ as before.  

To be able to use $\K$ as a parameter analogous to $k$ in ABJM theory, we rescale $x\equiv y/( 4\pi \K)$ and $p\equiv k / (2\pi)$. Under this rescaling, we have
 \es{rhohat3}{
  \hat \rho = e^{- U_\pm(\hat x)/2} e^{- T(\hat p)} e^{- U_\pm(\hat x)/2} \le({1\pm R\ov 2}\ri) \quad \to \quad    \hat \rho={1\ov 4\pi\K} e^{-U_\pm(\hat y)/2} e^{- T(\hat k)} e^{- U_\pm(\hat y)/2}\le({1\pm R\ov 2}\ri) \,,
 }
where we used that $U(\hat x)$ commutes with $R$, and for the $(+)$ sign
 \es{PSignUT}{
   U_+(y) = &\log \frac{\left( 4 \cosh^2 \le(y\ov4\K\ri) \right)^{N_f+c+1}\, \left( 4 \cosh^2  \le(y\ov2\K\ri)  \right)^{d-1/2}}{ \le(4\sinh^2  \le(y\ov4\K\ri) \ri)^{a}\le(4\sinh^2  \le(y\ov2\K\ri) \ri)^{b}}\ , \\
  T(k) = &\log \left(2 \cosh \le( k\ov 2\ri) \right) \, ,
 }
 while for the  $(-)$ sign
\es{MSignUT}{
   U_-(y) = &\log \frac{\left( 4 \cosh^2 \le(y\ov4\K\ri) \right)^{N_f+c}\, \left( 4 \cosh^2  \le(y\ov2\K\ri)  \right)^{d-1/2}}{ \le(4\sinh^2  \le(y\ov4\K\ri) \ri)^{a-1}\le(4\sinh^2  \le(y\ov2\K\ri) \ri)^{b}}\,,
 }
and $T(k)$ is as above.
 
After rescaling, we get the following nice properties:
\es{comasym}{
  \le[\hat y,\hat k\ri]&= 4\pi \K \, i \ ,\\
U_\pm(y) &\to {y\ov 2} \qquad (y\to \infty)\ , \\
 T(k) &\to {k\ov 2} \qquad (k\to \infty) \ .
 \qquad 
}
We then identify $\hbar=4\pi \K$, and calculate the area of the Fermi surface as a function of energy using the Wigner Hamiltonian~\eqref{WignerHamiltonian}. It is important to bear in mind that the projector halves the density of states, as consecutive energy eigenvalues correspond to eigenfunctions of opposite parity.\footnote{We can also think of the projection as  Neumann or Dirichlet boundary conditions at the origin.}\footnote{It was pointed out by~\cite{Assel:2015hsa} that halving the Fermi surface area of the Wigner Hamiltonian~\eqref{WignerHamiltonian} only counts the number of eigenvalues below energy $E$ to leading order in $\hbar$. From the more careful treatment of~\cite{Assel:2015hsa,Okuyama:2015auc} it turns out that this count misses a quarter of an eigenvalue.  We include this shift in our results. } To zeroth order, the phase space volume  enclosed by the Fermi surface is again given by~\eqref{phsp0}, and the correction is given by~\eqref{phsp1}.

The $\int_0^\infty dy \, U''(y)$ part of the latter formula seems to be problematic at first sight. For generic $a,b$ parameter values
\es{USmallDist}{
U(y)\sim \log \abs{y}\qquad (y\to 0) \ ,
}
and the integral is divergent. Physically, this divergence would be the consequence of the careless semi-classical treatment of a Fermi gas in a singular potential~\eqref{USmallDist}.  We will not have to deal with such subtleties, however, for the following reason.  In the cases of interest we either have $a+b=0$ or $a+b=1$---see Table~\ref{tab:Constants}. If $a+b=0$, we choose $U(y)$ of~\eqref{PSignUT} corresponding to the projection by $(1+R)/2$, which is regular at the origin. If $a+b=1$ we choose $U( y)$ of~\eqref{MSignUT} corresponding to the projection by $(1-R)/2$, and the potential is again regular.  With these choices, we can go ahead and calculate~\eqref{phsp1}.

For the number of eigenvalues below energy $E$ we get:
\es{phpstot}{
n(E)&={V\ov 4\pi \hbar} +{ {(-1)^{a+b}\ov 4}} ={V_0+\Delta V\ov 4\pi \hbar} +{ {(-1)^{a+b}\ov 4}} ={E^2 \ov 2\pi^2 \K}-{ \K+1-2d  \ov 8}+{ {(-1)^{a+b}\ov 4}}-{1\ov 24\K} \ ,
}
where the $4\pi \hbar$ in the denominator comes from the fact that there is one state per $2 \pi \hbar$ phase space volume and we are counting only ever other state.  The shift by $(-1)^{a+b}/4$ was obtained in~\cite{Assel:2015hsa,Okuyama:2015auc} from a more refined calculation of the number of eigenvalues.  Intuitively, we can  interpret the origin of this shift as follows: for $a+b=0$ we projected onto even, while for $a+b=1$ onto odd wave functions. Because the lowest eigenfunction is even, we have undercounted (overcounted) the number of eigenvalues by $1/4$ in the case of the even (odd) projection.\footnote{Because this argument implicitly assumes a uniform spacing of even and odd eigenvalues, it does not replace the rigorous calculations of~\cite{Assel:2015hsa,Okuyama:2015auc}. We thank Nadav Drukker for discussions of this issue.}     The constants $C$ and $B$ from~\eqref{JAsympMain} take the values
\es{CB}{
C&={1 \ov 2\pi^2 \K} \ , \qquad B=n_0+{\pi^2 C\ov 3}=-{ \K+1-2d \ov 8}+{ {(-1)^{a+b}\ov 4}}+{1\ov 8\K} \ .
}
Analogously to~\eqref{sp2nfree}, the  free energy $F$ has the following large $N$ expansion:
\es{sp2nfree}{
F&={2\sqrt{2}\, \pi \ov 3 }\,\K^{1/2}  \, N^{3/2} +\pi\le({ \K^{3/2}  \ov 4\sqrt{2}}+{\le(1-2d+{ {2 (-1)^{a+b}}}\ri)\,\K^{1/2}\ov  4\sqrt{2}}-{1\ov 4\sqrt{2} \,\K^{1/2}} \ri)\, N^{1/2} +\dots\ .
}
This result matches with the saddle point computation of Section~\ref{sec:LargeNUSp}---see~\eqref{FFinal3}.  As another check, note that the answer \eqref{sp2nfree} is invariant under the redefinition of parameters in \eqref{Shifts}, as should be the case.

\section{Discussion and outlook} \label{sec:Discussion}

We summarize the results obtained in this paper for the partition function of ${\cal N}=4$ gauge theories with classical gauge groups with matter consisting of one two-index tensor (anti)symmetric and $N_f$ fundamental hypermultiplets. The partition function takes the form
\es{JAsympMain2}{
Z(N)&={\cal A}(N_f) \, {\rm Ai}\le[C^{-1/3}(N-B)\ri]+ {\cal O}\le(e^{-\sqrt{N}}\ri)\ ,
}
where $C$ and $B$ are given in Table~\ref{tab:CBValues}. Using the relation $f_{3/2}=2/(3\sqrt{C})$ from~\eqref{FExpand}, we get agreement with the supergravity calculation~\eqref{Gota2}. 

From Table~\ref{tab:CBValues} one can see that there can be different field theories with the same AdS$_4 \times X$ dual.  For instance, the theories $O(2N)\,+A$, $O(2N+1)\,+A$, and $USp(2N)\,+S$ are all dual to AdS$_4 \times (S^7/\hat{D}_{N_f})_\text{free}$, whereas $O(2N)\,+S$ and $O(2N+1)\,+S$, as well as $USp(2N)\,+A$ with the shift $N_f\to N_f+4$, are all dual to AdS$_4 \times (S^7/{\hat D}_{N_f+2})$, where the orbifold action on $S^7$ is not free.   As mentioned in Section~\ref{sec:FieldTheories}, the supergravity backgrounds must be distinguished by discrete torsion flux of the three-form gauge potential.

More generally, the results collected in Table~\ref{tab:CBValues}, together with the results of~\cite{Marino:2011eh,Marino:2012az} for $U(N)$ quiver theories, represent predictions for M-theory computations that go beyond the leading two-derivative 11-d supergravity.  In the case of ABJM theory, the $k^{3/2}$ contribution to $f_{1/2}$ appearing~\eqref{FExpansion} is accounted for by the shift in the membrane charge from higher derivative corrections on the supergravity side~\cite{Bergman:2009zh}.  It would be very interesting to derive the shifts in membrane charge and to take into account higher derivative corrections on the supergravity side for the other examples.  Note that from the large $N$ expansion of the Airy function \eqref{JAsympMain2}, one obtains a universal logarithmic term in the free energy equal to $-\frac 14 \log N$;  this term matches a one-loop supergravity computation on AdS$_4 \times X$ \cite{Bhattacharyya:2012ye}.\footnote{We thank Nikolay Bobev for discussions on this issue.}  Perhaps one could derive the full Airy function behavior from supergravity calculations.

\begin{table}[!h]
\begin{center}
\begin{tabular}{c||c||c||c|c}
$G\,+$ matter & IIA orientifold & M-theory on AdS$_4\times X$ & $C$ & $B$\\
\hline
\hline
$U(N)\,+\text{adj}$ &no orientifold&$S^7 / \Z_{N_f}$& ${2 \ov \pi^2 N_f} $& $-{ N_f \ov 8}+{1\ov 2N_f}$\\
\hline
$O(2N)\,+A$ &O$2^-$&$(S^7/\hat{D}_{N_f})_\text{free}$& ${1 \ov 2\pi^2 N_f}$ & $-{ N_f-1 \ov 8}+{1\ov 8N_f}$\\
\hline
$O(2N)\,+S$  &O$6^+$&$S^7/{\hat D}_{N_f+2}$& ${1 \ov 2\pi^2 (N_f+2)}$ & $-{ N_f-1 \ov 8}+{1\ov 8(N_f+2)}$\\
\hline
$O(2N+1)\,+A$ &$\widetilde{\text{O}2}^-$&$(S^7/\hat{D}_{N_f})_\text{free}$ & ${1 \ov 2\pi^2 N_f}$ & $-{ N_f+3 \ov 8}+{1\ov 8N_f}$\\
\hline
$O(2N+1)\,+S$&O$6^+$&$S^7/{\hat D}_{N_f+2}$& ${1 \ov 2\pi^2 (N_f+2)}$ & $-{ N_f+3 \ov 8}+{1\ov 8(N_f+2)}$\\
\hline
$USp(2N)\,+A$&O$6^-$&$S^7/{\hat D}_{N_f-2}$& ${1 \ov 2\pi^2 (N_f-2)}$ & $-{ N_f+1 \ov 8}+{1\ov 8(N_f-2)}$\\
\hline
$USp(2N)\,+S$ &O$2^+$& $(S^7/\hat{D}_{N_f})_\text{free}$ & ${1 \ov 2\pi^2 N_f}$ & $-{ N_f+1 \ov 8}+{1\ov 8N_f}$\\
\end{tabular}
\end{center}
\caption{The values of the constants $C$ and $B$ appearing in~\eqref{JAsympMain2} for a gauge theory  with gauge group $G$, $N_f$ fundamental flavors, and a two-index antisymmetric $(A)$ or symmetric $(S)$ hypermultiplet. We also listed the type IIA construction, and dual M-theory geometry. To compare with the gravity calculation~\eqref{Gota2}, one needs the relation $f_{3/2}=2/(3\sqrt{C})$. The values of $B$ given in this table include the shift by $(-1)^{a+b}/4$ exhibited in~\eqref{CB} that was pointed out in~\cite{Assel:2015hsa,Okuyama:2015auc} after the original version of this paper appeared.}\label{tab:CBValues}
\end{table}%

It would be desirable to generalize the methods in this paper to more complicated quiver theories with classical gauge groups and Chern-Simons interactions. Although at first sight it may seem straightforward to generalize the large $N$ approximation of Section~\ref{sec:LargeN} to the more general setup, there are additional complications related to the non-smoothness of the eigenvalue distributions at leading order in large $N$ and the non-exact cancellation of long-range forces between eigenvalues at subleading order.  We leave such a general treatment for future work.  The Fermi gas approach explored in Section~\ref{sec:Fermi} is very powerful, but it relies crucially on non-trivial determinant formulae. It would be interesting to understand better the set of SCFTs with orthogonal and symplectic gauge groups that lend themselves to this approach.  One may hope that the $S^3$ partition functions of all theories with ${\cal N} \geq 3$ supersymmetry can be written as non-interacting Fermi gases.

\section*{Acknowledgments}
We thank O.~Aharony, O.~Bergman, N.~Bobev, D.~Freedman, A.~Hanany, M.~Mari\~no, S.-H.~Shao, and M.~Tierz for useful discussions.  This work was supported in part by the U.S. Department of Energy under cooperative research agreement Contract Number DE-FG02-05ER41360\@.  SSP was also supported in part by a Pappalardo Fellowship in Physics at MIT\@.

\appendix

\section{Quantum-corrected moduli space}
\label{MODULISPACE}

As a check that the field theories presented in Table~\ref{tab:Ingredients} are dual to M-theory on AdS$_4 \times X$, where $X$ is the quotient of $S^7$ in Table~\ref{tab:Ingredients}, one can make sure that the moduli space of these field theories does indeed match the moduli space of $N$ M2-branes probing the 11d geometry.    We will do so at the level of algebraic geometry, without explicitly constructing the full hyperk\"ahler metric on the moduli space.   In this computation, monopole operators play a crucial role, because they parameterize certain directions in the moduli space \cite{Gaiotto:2009tk,Jafferis:2009th}.  It is very important to include quantum corrections to their scaling dimensions, which essentially determine their OPE as in \cite{Gaiotto:2009tk,Jafferis:2009th,Benini:2009qs,Benini:2011cma}.

To define monopole operators, one should first consider monopole backgrounds.  We use the convention where for a gauge theory with gauge group $G$, the gauge field ${\cal A}$ corresponding to a GNO monopole background centered at the origin takes the form
 \es{GNO}{
  {\cal A} = H (\pm 1 - \cos \theta) d\phi \,,
 }
where $H$ is an element of the Lie algebra $\mathfrak{g}$.  Using the gauge symmetry, one can rotate $H$ into the Cartan $\{h_i\}$ subalgebra, namely
 \es{HRotate}{
  H = \sum_{i= 1}^r q_i h_i \,,
 }
where $r$ is the rank of $G$.  The Dirac quantization condition requires 
 \es{Dirac}{
  q \cdot w \in \Z/2
 } 
for any allowed weight $w$ of an irreducible representation of $G$.  These monopole backgrounds should be considered only modulo the action of the Weyl group.

The background \eqref{GNO} above breaks all supersymmetry by itself.  To define a supersymmetric background, one should supplement \eqref{GNO} with a non-trivial profile for one of the three real scalars in the ${\cal N} = 4$ vectormultiplet.  Let this scalar be $\sigma$;  we must take $\sigma = H / \abs{x}$.  The choice of such a scalar breaks the $SO(4)_R$ symmetry of the ${\cal N} = 4$ supersymmetry algebra to an $SO(2)_R$ subgroup corresponding to an ${\cal N} =2$ subalgebra.   In this ${\cal N} = 2$ language, one can define chiral monopole operators ${\cal M}_q$ corresponding to the GNO background described above.  Being chiral, one can identify their scaling dimension $\Delta_q$ with the $SO(2)_R$ charge.  As shown in \cite{Gaiotto:2008ak,Gaiotto:2009tk}, the BPS monopole operator ${\cal M}_q$ acquires at one-loop the R-charge 
 \es{RCharge}{
  \Delta_q = \sum_\text{hypers} \abs{q \cdot w} - \sum_\text{vectors} \abs{q \cdot w} \,,
 } 
where the sums run over all the weights $w$ of the fermions in the hyper and vectormultiplets.  As far as the ${\cal N} = 4$ supersymmetry is concerned, these chiral monopole operators ${\cal M}_q$ are the highest weight states of $SO(4)_R$ representations of dimension $2 \Delta_q + 1$.  However, only the chiral operators with scaling dimension \eqref{RCharge} will be relevant for us.

\subsection{${\cal N} = 4$ $U(N)$ gauge theory with adjoint and $N_f$ fundamental hypermultiplets}

Let us start by reviewing the construction of the geometric branch of the moduli space for the ${\cal N} = 4$ $U(N)$ gauge theory with an adjoint hyper and $N_f$ fundamental hypermultiplets.  In ${\cal N} = 2$ notation, the matter content of the theory consists of adjoint chiral multiplets with bottom components $X$, $Y$ (coming from the adjoint ${\cal N}=4$ hypermultiplet), and $Z$ (coming from the ${\cal N}=4$ vectormultiplet), as well as chiral multiplets with bottom components $q_i$, $i =1, \ldots, N_f$ transforming in the fundamental of $U(N)$ and $\widetilde q_i$ transforming in the conjugate representation. The ${\cal N} = 2$ superpotential, 
 \es{superpot}{
   W = \tr \left( Z[X, Y] + \widetilde q_i Z q_i  \right) \,, 
 }
is consistent with the R-charges of $X$, $Y$, $q_i$, and $\widetilde q_i$ being equal $1/2$, and that of $Z$ being equal $1$, as can be derived for instance using the $F$-maximization procedure \cite{Jafferis:2010un, Jafferis:2011zi, Closset:2012vg}.

In the $N=1$ case, the moduli space of this theory should match precisely the eight-dimensional transverse space probed by the M2-branes.  Indeed, in this case the chiral multiplets corresponding to $X$ and $Y$ completely decouple, and the expectation values of these complex fields parameterize a $\C^2$ factor in the moduli space of vacua.  The rest of the moduli space is parameterized by expectation values for $Z$ and for the monopole operators $T = {\cal M}_{1/2}$ and $\widetilde T = {\cal M}_{-1/2}$.  These operators are not independent;  in the chiral ring, they satisfy a relation that can be determined as follows.  According to \eqref{RCharge}, we can calculate their R-charge to be
 \es{DeltaUN}{
  \Delta = \frac{N_f}{2} + 0 - 0= \frac {N_f}{2} \,,
 }
where in the middle equality we exhibited explicitly the contributions from the $N_f$ fundamentals, the adjoints, and the ${\cal N} = 4$ vector, respectively.  One then expects the OPE \cite{Gaiotto:2009tk,Jafferis:2009th,Benini:2009qs,Benini:2011cma}
 \es{OPEUN}{
  T \widetilde T \sim Z^{N_f} \,,
 }
which should be imposed as a relation in the chiral ring.  Giving $T$,  $\widetilde T$, and $Z$ expectation values obeying \eqref{OPEUN} describes the orbifold $\C^2 / \Z_{N_f}$, as can be seen from ``solving'' \eqref{OPEUN} by writing $T = a^{N_f}$, $\widetilde T = b^{N_f}$, and $Z = ab$.  The coordinates $a$ and $b$ parameterize $\C^2 / \Z_{N_f}$ because both $(a, b)$ and $(a e^{2 \pi i / N_f}, b e^{-2 \pi i / N_f})$ yield the same point in \eqref{OPEUN}.  The moduli space of the $U(1)$ theory is therefore $\C^2 \times (\C^2 / \Z_{N_f})$, where the $\C^2$ factor is parameterized by the free fields $X$ and $Y$, and the $\C^2 / \Z_{N_f}$ factor is really just the complex surface \eqref{OPEUN}. Defining 
 \es{zDefs}{
  z_1 = X \,, \qquad z_2 = Y\,, \qquad z_3 = a \,, \qquad z_4 = b^* \,,
 }
we obtain the description of $\C^2 \times (\C^2 / \Z_{N_f})$ used around eq.~\eqref{ZNfAction}.

When $N>1$, the theory has a Coulomb branch where the fundamentals vanish and the adjoint fields $X$, $Y$, and $Z$ have diagonal expectation values
 \es{Diagonal}{
  X = \diag\{x_1, x_2, \ldots, x_N\} \,, \qquad
   Y = \diag\{y_1, y_2, \ldots, y_N\} \,, \qquad
    Z = \diag\{z_1, z_2, \ldots, z_N\}
 }
(to ensure vanishing of the F-term potential), thus breaking the gauge group generically to $U(1)^N$.  In addition, there are BPS monopole operators corresponding to
 \es{MonNonAb}{
  H = \diag\{q_1, q_2, \ldots, q_N \} \,;
 }
we can denote the BPS monopole operators with $q_i = \pm 1/2$ and $q_j = 0$ with $i \neq j$ by $T_i$ (for the plus sign) and $\widetilde T_i$ (for the minus sign).  An argument like the one in the Abelian case above shows that for every $i$, we have
 \es{OPENonAb}{
  T_i \widetilde T_i \sim z_i^{N_f} \,.
 }
For each $i$ we therefore have a $\C^2 \times (\C^2 / \Z_{N_f})$ factor in the moduli space parameterized by $(x_i, y_i, z_i, T_i, \widetilde T_i)$.  The Weyl group of $U(N)$ acts by permuting these factors, so the Coulomb branch of the $U(N)$ theory is the symmetric product of $N$ $\C^2 \times (\C^2 / \Z_{N_f})$ factors.  This space is precisely the expected Coulomb branch of $N$ M2-branes probing $\C^2 \times (\C^2 / \Z_{N_f})$.  In addition to the Coulomb branch, the moduli space also has a Higgs branch where the fundamental fields $q$ and $\widetilde q$ acquire expectation values.  This branch is not realized geometrically, however, and is therefore of no interest to us.

\subsection{The $USp(2N)$ theories}

Let us now consider the ${\cal N} = 4$ $USp(2N)$ theories with $N_f$ fundamental and either a symmetric or an anti-symmetric hypermultiplet.  As in the $U(N)$ case, let us denote the bottom components of the ${\cal N} =2 $ matter multiplets by $X$, $Y$ (transforming either in the symmetric or anti-symmetric representations of $USp(2N)$), $Z$ (transforming in the adjoint (symmetric) representation of $USp(2N)$), $q_i$, and $\widetilde q_i$ (transforming in the fundamental representation).  A superpotential like \eqref{superpot} determines the R-charges of these operators just like in the $U(N)$ case.

In the $N=1$ case, we expect that the Coulomb branch of the $USp(2) \cong SU(2)$ theory should match the geometry probed by the M2-branes.  In this case, $Z$ is an adjoint field, while $X$ and $Y$ are either adjoint-valued or singlets under $SU(2)$ (corresponding to symmetric or anti-symmetric $USp(2N)$ tensors, respectively).   On the Coulomb branch, the gauge group is broken to $U(1)$ by expectation values for the adjoint fields.  Without loss of generality, we can consider these expectation values to be in the $J_3 =  \frac 12 \sigma_3$ direction and take $Z = z J_3$.  If $X$ and $Y$ are adjoint-valued, we should also take $X = x J_3$, $Y = y J_3$ (such that the bosonic potential following from the first term in \eqref{superpot} vanishes);   if $X$ and $Y$ are $SU(2)$ singlets, we can take $X = x$ and $Y = y$.  The expectation values of the fundamental fields $q_i$ and $\widetilde q_i$ must vanish for any of the vacua on the Coulomb branch. 

As in the $U(1)$ case, a full description of the moduli space also involves monopole operators.  For an $SU(2)$ gauge theory, the monopole operators can be taken to correspond to a GNO background \eqref{GNO} with
 \es{HSU2}{
  H = q J_3 \,.
 }
Since the possible weights $w$ are half-integral, the Dirac quantization condition \eqref{Dirac} implies $q \in \Z$.  Note that all these operators are topologically trivial, because the group $SU(2)$ has trivial topology. The operators with smallest charge are $T = {\cal M}_1$ and $\widetilde T = {\cal M}_{-1}$.  In fact, in the $SU(2)$ gauge theory, $T$ and $\widetilde T$ are identified under the action of the Weyl group, but on the Coulomb branch they will be distinct.  According to \eqref{RCharge}, the R-charge of $T$ and $\widetilde T$ is 
 \es{RchargeSU2}{
  \Delta = \begin{cases} N_f + 2 - 2 = N_f & \text{if $X$, $Y$ are adjoints} \,, \\
    N_f + 0 - 2 = N_f - 2& \text{if $X$, $Y$ are singlets} \,,
   \end{cases} 
 }
where, as in \eqref{DeltaUN}, in the middle equality we exhibited explicitly the contributions from the $N_f$ fundamental, from the adjoints / singlets, and from the ${\cal N} = 4$ vector, respectively.  From \eqref{RchargeSU2}, one expects the OPE
 \es{OPESU2}{
  T \widetilde T \sim \tr Z^{2 \Delta} = (\tr Z^2)^{\Delta} \,,
 } 
where the trace is taken in the fundamental representation of $SU(2)$. The relation \eqref{OPESU2} should be imposed as a relation in the chiral ring.  Note that $\Delta$ in \eqref{RchargeSU2} is always an integer, so $\tr Z^{2 \Delta}$ does not vanish.  Also note that if $N_f=0$ in the adjoint case and $N_f \leq 2$ in the singlet case we obtain monopole operators with R-charge $\Delta \leq 0$, which signifies that one of the assumptions in our UV description of the theory must break down as we flow to the IR critical point.  Such theories were called ``bad'' in \cite{Gaiotto:2008ak}, and we will not examine them.   See also footnotes~\ref{Nf0Footnote} and~\ref{Nf2Footnote}.

We are now ready to give the full description of the Coulomb branch.  It is parameterized by the complex fields $x$, $y$, $z$, $T$, and $\widetilde T$.  The latter three satisfy
 \es{OPESU2Coulomb}{
  T \widetilde T \sim z^{2 \Delta} \,,
 }
as can be easily seen from \eqref{OPESU2}.  In addition, $SU(2)$ has a $\Z_2$ Weyl group, which sends $J_3 \to -J_3$ and consequently also acts non-trivially on the Coulomb branch by flipping the sign of the adjoint fields and interchanging $T$ with $\widetilde T$:
 \es{WeylAction}{
  \text{$X$, $Y$ adjoints (symmetric):} &\qquad 
   (x, y, z, T, \widetilde T) \sim (-x, -y, -z, \widetilde T, T) \,, \\
  \text{$X$, $Y$ singlets (anti-symmetric):} &\qquad 
   (x, y, z, T, \widetilde T) \sim (x, y, -z, \widetilde T, T) \,.
 }
We can relate this description of the Coulomb branch to the one used in Section~\ref{BRANE}.  First, we solve \eqref{OPESU2Coulomb} by writing $T = a^{2\Delta}$, $\widetilde T = (c^*)^{2\Delta}$, and $z = a c^*$, for some complex coordinates $a$ and $c$.  There is a redundancy in this description that forces us to make the identification
 \es{acIdentif}{
  (a, c) \sim e^{\pi i / \Delta} (a, c) \,.
 }
In terms of $a$ and $c$, the Weyl group identifications \eqref{WeylAction} yield
 \es{WeylActionAgain}{
  \text{$X$, $Y$ adjoints (symmetric):} &\qquad 
   (x, y, a, c) \sim (-x, -y, i c^*, -i a^*) \,, \\
  \text{$X$, $Y$ singlets (anti-symmetric):} &\qquad 
   (x, y, a, c) \sim (x, y, i c^*, -i a^*) \,.
 }
Redefining $x = z_1$, $y = z_2$, $a = z_3$, $c = z_4$, we obtain precisely the orbifold description used in Section~\ref{BRANE}.  In the case where $X$ and $Y$ are adjoint fields, the Coulomb branch is a freely acting $\hat D_{N_f}$ orbifold of $\C^4$, while in the case where $X$ and $Y$ are singlets the Coulomb branch is a $\hat D_{N_f-2}$ orbifold of $\C^4$ (more precisely $\C^2 \times (\C^2 / \hat D_{N_f-2})$) that now has fixed points because the coordinates $z_1 = x$ and $z_2 = y$ are invariant under the action \eqref{acIdentif}--\eqref{WeylActionAgain}.

When $N>1$, the moduli space of vacua has a Coulomb branch where the fundamentals vanish and $X$, $Y$, and $Z$ acquire expectation values and break $USp(2N)$ generically to $U(1)^N$.  A straightforward analysis shows that if $X$ and $Y$ are symmetric tensors, the Coulomb branch is the $N$th symmetric power of the space $\C^4 / \hat D_{N_f}$ found above in the $N=1$ case;  if $X$ and $Y$ are anti-symmetric tensors, the Coulomb branch is the $N$th symmetric power of $\C^2 \times (\C^2 / \hat D_{N_f-2})$.  These spaces are precisely the expected moduli spaces of $N$ M2-branes probing $\C^4 / \hat D_{N_f}$ or $\C^2 \times (\C^2 / \hat D_{N_f-2})$.  In addition to the Coulomb branch, the theory also has a Higgs branch where the fundamentals $q_i$ and $\widetilde q_i$ acquire VEVs, but the Higgs branch is not realized geometrically in M-theory.

It is worth noting that if $X$ and $Y$ are anti-symmetric tensors of $USp(2N)$, one could consider imposing a symplectic tracelessness condition on these fields.  When $N=1$, the fields $X$ and $Y$ would be completely absent, because what survived in the analysis above was precisely their symplectic trace.  The moduli space would therefore be only $\C^2 / \hat D_{N_f}$ if the symplectic trace were removed from $X$ and $Y$, and it would not match the eleven-dimensional geometry.  The correct field theory that arises from the brane construction of  Section~\ref{BRANE} is that where $X$ and $Y$ are not required to be symplectic traceless.

\subsection{The $O(2N)$ theories}

The discussion of the Coulomb branch for ${\cal N} = 4$ $O(2N)$ gauge theory with either a symmetric or anti-symmetric hypermultiplet and $N_f$ hypermultiplets in the the fundamental representation of $O(2N)$ parallels the discussion of the $USp(2N)$ case above.  Let $X$, $Y$, $Z$, $q_i$, and $\widetilde q_i$ be the bottom components of the chiral multiplets arising from the ${\cal N} = 4$ hyper and vectormultiplets as before.  Now $Z$ transforms in the adjoint (anti-symmetric) representation of $O(2N)$, while $X$ and $Y$ transform either in the symmetric or antisymmetric tensor representations of $O(2N)$.  While the representations of $O(2N)$ are real, and therefore there exists the possibility of considering real scalar fields, our scalar fields $X$, $Y$, $Z$, $q_i$, and $\widetilde q_i$ are all complex.

In the case $N=1$, one can again study the Coulomb branch where the fundamentals all vanish and $X$, $Y$, and $Z$ have expectation values.  Let 
 \es{JSO2}{
  J = \begin{pmatrix} 0 & -i \\
  i & 0 
  \end{pmatrix}
 }
be the Hermitian generator of $O(2)$.  On the Coulomb branch, we should take $Z = z J$.  If $X$ and $Y$ are symmetric matrices, the scalar potential vanishes if these matrices commute with $Z$, so we should take $X = x I_2$ and $Y = y I_2$, where $I_2$ is the $2 \times 2$ identity matrix.  If $X$ and $Y$ are anti-symmetric matrices, the only option is $X = x J$ and $Y = y J$ for some complex numbers $x$ and $y$.

The monopole operators for an $O(2)$ gauge theory correspond to backgrounds \eqref{GNO} with 
 \es{HSO2}{
  H = q J \,.
 }
Since the possible weights of $O(2)$ representations are all integral, Dirac quantization implies $q \in \Z/2$.  The monopole operators of smallest charge are $T = {\cal M}_{1/2}$ and $\widetilde T = {\cal M}_{-1/2}$.  These operators are independent at generic points on the Coulomb branch.  If the gauge group were $SO(2)$ they would also be distinct at the CFT fixed point at the origin of the Coulomb branch, but for an $O(2)$ gauge group they get identified:  Indeed, one can consider the charge conjugation
 \es{parity}{
  C = \begin{pmatrix}
   1 & 0 \\
   0 & -1
  \end{pmatrix} \in O(2) \,,
 }
(which in the $O(2)$ theory is gauged) under which $J \to C J C^{-1} = -J$.  Charge conjugation identifies $T$ with $\widetilde T$ at the origin of moduli space because it identifies the defining backgrounds \eqref{JSO2}.  The R-charge of $T$ and $\widetilde T$ is
 \es{RchargeO2}{
  \Delta = \begin{cases} N_f + 2 - 0 = N_f + 2 & \text{if $X$, $Y$ are symmetric} \,, \\
    N_f + 0 - 0 = N_f & \text{if $X$, $Y$ are anti-symmetric} \,.
   \end{cases} 
 }
These R-charges imply that $T$ and $\widetilde T$ satisfy the OPE \eqref{OPESU2}, as in the $SU(2)$ case.

The Coulomb branch is parameterized by the complex parameters $x$, $y$, $z$, $T$, and $\widetilde T$ satisfying the constraint \eqref{OPESU2Coulomb}.  Charge conjugation \eqref{parity} imposes the further identifications
 \es{WeylActionO2}{
  \text{$X$, $Y$ symmetric:} &\qquad 
   (x, y, z, T, \widetilde T) \sim (x, y, -z, \widetilde T, T) \,, \\
  \text{$X$, $Y$ anti-symmetric:} &\qquad 
   (x, y, z, T, \widetilde T) \sim (-x, -y, -z, \widetilde T, T) \,.
 }
Writing $T = a^{2\Delta}$, $\widetilde T = (c^*)^{2\Delta}$, and $z = a c^*$ as in the $SU(2)$ case, we obtain a description of the Coulomb branch in terms of the complex parameters $(x, y, a, c)$ subject to the identifications $(a, c) \sim e^{\pi i / \Delta}(a, c)$ and 
 \es{WeylActionAgainO2}{
  \text{$X$, $Y$ symmetric:} &\qquad 
   (x, y, a, c) \sim (x, y, -z, i c^*, -i a^*) \,, \\
  \text{$X$, $Y$ anti-symmetric:} &\qquad 
   (x, y, a, c) \sim (-x, -y, i c^*, -i a^*) \,.
 }
Denoting $x = z_1$, $y = z_2$, $a = z_3$, and $c = z_4$ as in the $SU(2)$ case we obtain the same description of the eight-dimensional hyperk\"ahler space that appears in the eleven-dimensional geometry, as described in Section~\ref{BRANE}.  

In the $N>1$ case, one can check that the Coulomb branch is the $N$th symmetric power of $\C^2 \times (\C^2 / \hat D_{N_f})$ or $\C^4 / \hat D_{N_f}$, depending on whether $X$ and $Y$ are symmetric or anti-symmetric tensors, respectively.  This geometry matches precisely the moduli space of $N$ M2-branes probing, respectively, $\C^2 \times (\C^2 / \hat D_{N_f})$ or $\C^4 / \hat D_{N_f}$.  As in the $U(N)$ and $USp(2N)$ theories discussed above, there is also a Higgs branch where the fundamentals have expectation values, but this branch is not realized geometrically in M-theory.

Note that having an $O(2N)$ gauge group as opposed to $SO(2N)$ was very important for matching the eleven-dimensional geometry.  In an $SO(2)$ gauge theory, the identifications~\eqref{WeylActionO2} and \eqref{WeylActionAgainO2} would not be present.  Note also that in the case where $X$ and $Y$ are symmetric tensors of $O(2N)$, one has in principle the possibility of imposing a tracelessness condition on $X$ and $Y$.  When $N=1$, the moduli space would then be $\C^2 / \hat D_{N_f}$, and would therefore have complex dimension two.  For the field theory that arises from the orientifold construction in string theory one should therefore not require $X$ and $Y$ to be traceless.

\subsection{The $O(2N+1)$ theories}

Lastly, let us consider the ${\cal N} = 4$ $O(2N+1)$ gauge theories with either a symmetric or anti-symmetric tensor hypermultiplet and $N_f$ fundamental hypermultiplets.  We use the same notation for the bottom components of the various ${\cal N} = 2$ chiral multiplets as in the previous section.

In the $N=1$ case, we can take the theory to the Coulomb branch by giving an expectation value to $Z = z J_{12}$ to the complex scalar $Z$ belonging to the ${\cal N} = 4$ vectormultiplet.  Here,
 \es{J12Def}{
  J_{12} = \begin{pmatrix}
   0 & -i & 0 \\
   i & 0 & 0 \\
   0 & 0 & 0 
  \end{pmatrix}
 }
is the generator of rotations in the $12$-plane in color space.  To ensure that the scalar potential vanishes, one should also take $X = x J_{12}$ and $Y = y J_{12}$ in the case where $X$ and $Y$ are anti-symmetric tensors, and $X = \diag\{x, x, \widetilde x \}$, $Y = \diag\{y, y, \widetilde y\}$ in the case where $X$ and $Y$ are symmetric tensors.  In both cases, the vanishing of the F-term potential requires $q_i = \widetilde q_i = 0$.  

The relevant BPS monopole operators in this case correspond to
 \es{MonO2N1}{
  H = q J_{12} \,.
 } 
Dirac quantization implies $q \in \Z/2$, and as before we denote $T = {\cal M}_{1/2}$ and $\widetilde T = {\cal M}_{-1/2}$.  The operators $T$ and $\widetilde T$ are distinct on the Coulomb branch, but at the CFT fixed point they get identified.  Indeed, the gauge transformations corresponding to
 \es{gaugeTransf}{
  O = \begin{pmatrix} 1 & 0 & 0 \\
  0 & -1 & 0 \\
  0 & 0 & \pm 1 
  \end{pmatrix} \in O(3)
 } 
send $J_{12} \to O J_{12} O^{-1} = -J_{12}$, so they identify $T$ with $\widetilde T$.  Note that the minus sign in \eqref{gaugeTransf} yields a group element of $SO(3)$ as well, while the plus sign does not;  the transformation corresponding to the plus sign is the charge conjugation symmetry of the $SO(3)$ theory, which is gauged when the gauge group is $O(3)$.  Unlike the $O(2)$ case discussed above, the operators $T$ and $\widetilde T$ are not identified only in the $O(3)$ gauge theory.  They would be identified in the $SO(3)$ gauge theory as well.  The R-charge of $T$ and $\widetilde T$ is
 \es{RchargeO3}{
  \Delta = \begin{cases} N_f + 3 - 1 = N_f + 2 & \text{if $X$, $Y$ are symmetric} \,, \\
    N_f + 1 - 1 = N_f & \text{if $X$, $Y$ are anti-symmetric} \,.
   \end{cases} 
 }
Based on these R-charges, one can infer that $T$ and $\widetilde T$ satisfy the OPE \eqref{OPESU2}.

The Coulomb branch in this case is parameterized by $x$, $y$, $z$, $T$, $\widetilde T$, as well as $\widetilde x$ and $\widetilde y$ in the symmetric tensor case.  The fields $z$, $T$, and $\widetilde T$ satisfy the chiral ring relation \eqref{OPESU2Coulomb}.  The transformation \eqref{gaugeTransf} imposes the same relations as in \eqref{WeylActionO2}--\eqref{WeylActionAgainO2} and does not act on $\widetilde x$ and $\widetilde y$.  The Coulomb branch is therefore $\C^4 / \hat D_{N_f+2}$ if $X$ and $Y$ are anti-symmetric tensors, just like in the $O(2)$ case discussed above.  If $X$ and $Y$ are symmetric tensors, the Coulomb branch is $\C^2 \times \C^2 \times \C^2 / \hat D_{N_f + 2}$, where the extra $\C^2$ factor relative to the $O(2)$ case is parameterized by $\widetilde x$ and $\widetilde y$.

This discussion generalizes to $N>1$. If $X$ and $Y$ are anti-symmetric tensors, the Coulomb branch is the $N$th symmetric power of $\C^4 / \hat D_{N_f+2}$, as expected from $N$ M2-branes probing $\C^4 / \hat D_{N_f}$.  If $X$ and $Y$ are symmetric tensors, the Coulomb branch is $\C^2$ times the $N$th symmetric power of $\C^2 \times \C^2 / \hat D_{N_f+2}$.  This moduli spaces is also as expected from $N$ M2-branes probing $\C^2 \times (\C^2 / \hat D_{N_f})$, together with a fractional M2-brane that is stuck at the $\C^2 / \hat D_{N_f}$ singularity and can only explore the $\C^2$ part of the geometry.  This fractional M2-brane corresponds to the half-D2-brane that is stuck to the O$6^+$-plane.  As in the previous cases, the moduli space also has a Higgs branch where the fundamental fields $q_i$ and $\widetilde q_i$ have expectation values, but this branch is not realized geometrically.

\section{Lightning review of the Fermi gas method of~\cite{Marino:2011eh}}  \label{app:FermiGas}

In this Appendix we review briefly the approach of~\cite{Marino:2011eh} for computing the partition function of a non-interacting Fermi gas.  For such a system, the number of energy eigenvalues below some energy $E$ is given by:
\es{NumberEigen}{
n(E)=\Tr \theta(E-\hat{H})=\sum_n \theta(E-E_n) \ ,
}
where $E_n$ is the $n$th energy eigenvalue of the full system. The density of states is defined by
\es{DoS}{
\rho(E)={dn(E)\ov dE}=\sum_n \delta(E-E_n) \ .
}
In the thermodynamic limit, $\rho(E)$ becomes a continuous function. The grand canonical potential of the non-interacting gas is given by:
\es{GrandPotential}{
J(\mu)=\int_0^\infty dE\ \rho(E)\log\le(1+e^{-E+\mu}\ri) \,,
}
where $\mu$ is the chemical potential. The canonical partition function and the free energy can be obtained from evaluating
\es{FreeGrand}{
Z(N)&={1\ov 2\pi i}\,\int d\mu \ e^{J(\mu)-\mu N} \,, \\
F(N)&=-\log Z(N) \ .
}

In the thermodynamic limit where $N\to\infty$, we only need the asymptotic form of $n(E)$ in order to determine the free energy to all orders in $1/N$. In the models of interest, we find that
\es{nAsymp}{
n(E)=C\,E^2+n_0+{\cal O}\le(E\, e^{-E}\ri) \ .
}
Then a short calculation gives
\es{JAsymp}{
J(\mu)&={C\ov 3}\, \mu^3+B\, \mu+A+{\cal O}\le(\mu\, e^{-\mu}\ri)\ , \qquad B=n_0+{\pi^2\,C\ov 3} \ , \\
Z(N)&={\cal A} \, {\rm Ai}\le[C^{-1/3}(N-B)\ri]+ {\cal O}\le(e^{-\sqrt{N}}\ri) \ , 
}
where $A$ and ${\cal A}=C^{-1/3} e^A$ are $N$-independent constants. 

The constants $C$ and $n_0$ can be determined by semiclassical methods, as they describe the large energy behavior of the density of states of the non-interacting Fermi gas. The semiclassical computation can be performed in the Wigner--Kirkwood formalism. Firstly, we introduce the Wigner transform of an operator $\hat{A}$:
\es{WignerTransform}{
A_W(x,p)=\int dy\ \bra{x-{y\ov2}} \hat{A} \ket{x+{y\ov2}}\, e^{ipy/\hbar} \ .
}
The Wigner transform obeys the product rule
\es{StarProduct}{
\le(\hat{A} \hat{B}\ri)_W&=A_W\star B_W \ , \\
\star&\equiv\exp\le[{i\hbar\ov 2}\le(\overleftarrow{\p_x}\, \overrightarrow{\p_p}-\overleftarrow{\p_p}\, \overrightarrow{\p_x}\ri)\ri] \ .
}
The trace of an operator is given by the phase space integral of the Wigner transform:
\es{WignerTrace}{
\Tr \hat{A}=\int {dx\,dp\ov 2\pi\,\hbar}\, A_W(x,p) \ .
}

In the Fermi gases of interest in this paper, we will encounter one particle density matrices of the form\footnote{We have additional an projection operator multiplying this density matrix. We discuss the consequences of the projector in the main text.}
 \es{rhohatApp}{
  \hat \rho = e^{- U(\hat x)/2} e^{- T(\hat p)} e^{- U(\hat x)/2} \,,
 }
where $U(x)$ and $T(p)$ approach linear functions exponentially fast for large $x$ or $p$. One can then calculate the Wigner Hamiltonian
\es{WignerHamiltonian}{
\rho_W&\equiv e_\star^{-H_W}\ , \\
H_W(x,p)&=T(p)+U(x)+{\hbar^2\ov 24}\le[U'(x)^2\, T''(p)-2T'(p)^2\, U''(x)\ri] +{\cal O}(\hbar^4) \ .
}
Combining~\eqref{NumberEigen} and~\eqref{WignerTrace} we get
\es{PhaseSpace}{
n(E)=\int {dx\,dp\ov 2\pi\,\hbar}\  \theta(E-\hat{H})_W(x,p) \ .
}
Along the lines of the argument in~\cite{Marino:2011eh}, one can show that up to exponentially small corrections $n(E)$ is given by the phase space area 
\es{PhaseSpace2}{
n(E)=\int_{H_W^{(2)}(x,p)\leq E} {dx\, dp \ov 2\pi\,\hbar} +{\cal O}\le(E\, e^{-E}\ri) \  ,
}
where $H_W^{(2)}$ is the Wigner Hamiltonian through ${\cal O}(\hbar^4)$ displayed in~\eqref{WignerHamiltonian}. $n(E)$ is just the Fermi surface area the non-interacting Fermi gas fills up. \eqref{PhaseSpace2} can be evaluated using the explicit form of $U(x)$ and $T(p)$.

\section{Derivation of~\eqref{MomRepCosh}} \label{app:Fourier}

Let us note that using simple trigonometric identities
\es{TrigIdent}{
  \frac{4 \cosh (\pi x_1) \cosh(\pi x_2) }{4 \cosh \left( \pi (x_1 - x_2) \right) \cosh \left( \pi (x_1 + x_2) \right)}
   =  {1\ov  2\cosh \left( \pi (x_1 - x_2) \right) }+ {1\ov  2\cosh \left( \pi (x_1 + x_2) \right) }\,.
 }
Using the Fourier transform
\es{FourierIdent}{
\int dx \ e^{2 \pi i p x } \, {1\ov \cosh \pi p}=\frac{1}{\cosh \pi x} \ ,
}
it is easy to see that
\es{MomRepTerm}{
 {1\ov  2\cosh \left( \pi (x_1 - x_2) \right) }= \left\langle x_1 \Bigg\vert \frac{1 }{2 \cosh (\pi \hat p)}  \Bigg\vert  x_2 \right\rangle \ ,
}
where in $h=1$ units $\hat{p}=-{1\ov 2\pi i}\, \p_x$ in position space.  Finally, we make use of $R\ket{x_2}=\ket{-x_2}$ to combine~\eqref{TrigIdent} and~\eqref{MomRepTerm}:
 \es{MomRepCoshApp}{
  \frac{4 \cosh (\pi x_1) \cosh(\pi x_2) }{4 \cosh \left( \pi (x_1 - x_2) \right) \cosh \left( \pi (x_1 + x_2) \right)}
   = \left\langle x_1 \Bigg\vert \frac{1 + R}{2 \cosh (\pi \hat p)}  \Bigg\vert  x_2 \right\rangle \, .
 }
This expression is the same as~\eqref{MomRepCosh}, which is what we set out to show.

\section{Derivation of the determinant formula}  \label{app:Determinant}

The Cauchy determinant formula states that for any numbers $u_i$ and $v_i$, with $1 \leq i, j \leq N$, the following identity holds
 \es{Cauchy}{
  \frac{\prod_{i<j} (u_i - u_j) (v_i - v_j) }{\prod_{i, j} (u_i + v_j)} = \det \frac{1}{u_i + v_j} \,.
 }
In this Appendix we derive a similar determinant formula\footnote{This formula has previously appeared in the literature.  See for example \cite{2000math......8184K}.  We thank Miguel Tierz for pointing this fact out to us.}:
 \es{CauchyGen}{
  \frac{\prod_{i<j} (u_i - u_j) (v_i - v_j) (u_i u_j - 1) (v_i v_j - 1)}{\prod_{i, j} (u_i + v_j) (u_i v_j + 1)} = \det \frac{1}{(u_i + v_j)(u_i v_j + 1)} \,.
 }

The proof of \eqref{CauchyGen} is very similar to the proof of the Cauchy formula \eqref{Cauchy}.   Let us compute the determinant of the matrix $M^{(N)}$ where
 \es{Mentries}{
  M^{(N)}_{ij} = \frac{1}{(u_i + v_j) (u_i + v_j^{-1})} \,.
 }
Subtracting the last column from each column $j <N$, one obtains the following entries
\es{Difference}{
 \frac{1}{(u_i + v_j)(u_i + v_j^{-1})} - \frac{1}{(u_i + v_N)(u_i  + v_N^{-1})} = 
  \frac{u_i(v_N - v_j) (v_N - v_j^{-1}) }{v_N(u_i + v_N)(u_i  + v_N^{-1})} \times\frac{1}{(u_i + v_j)(u_i + v_j^{-1})}  \ .
}
Extracting a factor of $1/(u_i + v_N)(u_i + v_N^{-1})$ from each row $i$ and a factor of $(v_N - v_j) (v_N - v_j^{-1}) / v_N$ from each column $j <N$, one obtains
 \es{MDet}{
  \det M^{(N)} = \frac{\prod_{j = 1}^{N-1}(v_N - v_j) (v_N - v_j^{-1})  }
   {v_N^{N-1} \prod_{i=1}^N (u_i + v_N)(u_i + v_N^{-1}) } 
   \begin{vmatrix}
     \frac{u_1}{(u_1 + v_1)(u_1 + v_1^{-1})}   & \ldots & \frac{u_1}{(u_1 + v_N)(u_1 + v_N^{-1})} & 1\\
     \frac{u_2}{(u_2 + v_1)(u_2 + v_1^{-1})}  & \ldots & \frac{u_2}{(u_2 + v_N)(u_2 + v_N^{-1})} & 1 \\
    \vdots & \vdots & \ddots & \vdots \\
     \frac{u_N}{(u_N + v_1)(u_N + v_1^{-1})} & \ldots & \frac{u_N}{(u_N + v_N)(u_N + v_N^{-1})} & 1
   \end{vmatrix} \,.
 }
In the determinant in \eqref{MDet} we now subtract the last row from each row $i <N$.  The entries for $i, j < N$ become
 \es{Difference2}{
  \frac{u_i}{(u_i + v_j)(u_i + v_j^{-1})} -  \frac{u_N}{(u_N + v_j)(u_N + v_j^{-1})}
   = \frac{(u_N - u_i) (u_N u_i - 1)}{(u_N + v_j)(u_N + v_j^{-1})} \frac{1}{(u_i + v_j)(u_i + v_j^{-1})} \,.
 }
Extracting a factor of $(u_N - u_i) (u_N u_i - 1)$ from each row $i < N$ and a factor of $1/(u_N + v_j)(u_N + v_j^{-1})$ from each column $j < N$ one obtains
 \es{MDetRecursion}{
  \det M^{(N)} =  \frac{\prod_{i=1}^{N-1} (u_N - u_i) (u_N u_i - 1) \prod_{j = 1}^{N-1}(v_N - v_j) (v_N - v_j^{-1})  }
   {v_N^{N-1} \prod_{i=1}^N (u_i + v_N)(u_i + v_N^{-1})  \prod_{j=1}^{N-1} (u_N + v_j)(u_N + v_j^{-1}) } \det M^{(N-1)} \,.
 }

By induction, we can then show
 \es{detM}{
  \det M^{(N)} = \frac{\prod_{j>i} (u_j - u_i) (u_i u_j -1) (v_j - v_i) (v_j - v_i^{-1})}
   {\prod_{i = 1}^N v_i^{i-1} \prod_{i, j} (u_i + v_j)(u_i + v_j^{-1})} \,.
 }
Since
 \es{detCG}{
  \det \frac{1}{(u_i + v_j) (u_i v_j + 1)} = \frac{1}{\prod_{i=1}^N v_i } \det M^{(N)} \,,
 }
after rearranging the factors of $v_i$ in \eqref{detM} we have
 \es{detCGAgain}{
  \det \frac{1}{(u_i + v_j) (u_i v_j + 1)} 
   =  \frac{\prod_{j>i} (u_j - u_i) (u_i u_j -1) (v_j - v_i) (v_i v_j -1)}
   { \prod_{i, j} (u_i + v_j)(u_i v_j + 1)} \,,
 }
which is the same expression as \eqref{CauchyGen}.

\bibliographystyle{ssg}
\bibliography{USp}

\end{document}